\documentclass[aps,prd,superscriptaddress]{revtex4-2}
\usepackage{graphicx}
\usepackage{color,amsmath}
\usepackage{here}
\usepackage[hang,small,bf]{caption}
\usepackage{wrapfig}
\usepackage{subcaption}
\usepackage{bbm}
\usepackage{slashed}
\usepackage{hyperref}
\usepackage[nameinlink]{cleveref}
\usepackage{xifthen}
\usepackage{xcolor}
\hypersetup{
	colorlinks,
	linkcolor={red!75!black},
	citecolor={blue!75!black},
	urlcolor={blue!75!black}
}

\newcommand{\Slash}[1]{\ooalign{\hfil/\hfil\crcr$#1$}}
\newcommand{\pa}{\partial}
\newcommand{\nn}{\nonumber}

\newcommand{\ep}{\epsilon}

\newcommand{\al}{\alpha}

\begin{document}
\title{Functional renormalization group flows and gauge consistency in QED}
\author{Yoshio Echigo}
\affiliation{Niigata Polytechnic College, Shibata 957-0017, Japan}

\author{Yuji Igarashi}
\email{igarashi@ed.niigata-u.ac.jp}
\affiliation{Faculty of Education, Niigata University, Niigata 950-2181, Japan}

\author{Katsumi Itoh}
\email{itoh@ed.niigata-u.ac.jp}
\affiliation{Faculty of Education, Niigata University, Niigata 950-2181, Japan}

\author{Jan M. Pawlowski}
\email{j.pawlowski@thphys.uni-heidelberg.de}
\affiliation{Institut f\"ur Theoretische Physik, Universit\"at Heidelberg, 
Philosophenweg 16, D-69120 Heidelberg, Germany }

\author{Yu Takahashi}
\email{f15j004a@alumni.niigata-u.ac.jp}
\affiliation{1196 Sakura-bayashi, Nishikan-ku, Niigata 953-0064, Japan }

\begin{abstract}
We consider quantum electrodynamics with chiral four-Fermi 
interactions in the functional renormalization group approach. In gauge
theories, the functional flow equation for the effective action is
accompanied by the quantum master equation that governs the underlying
gauge symmetry. Beyond perturbation theory, fully gauge-consistent
solutions are very difficult to obtain. We devise a systematic expansion
scheme in which the solutions of the flow equation also solve the
Quantum Master Equation. In the present work we apply this construction
within the lowest order corrections in the photon two-point
functions. In this truncation we discuss the phase structure in terms of
the gauge and four-Fermi couplings based on a numerical solution of the
system.
\end{abstract}

\maketitle

\newpage
\tableofcontents

\section{Introduction}
\label{sec:Introduction}

The functional renormalization group (fRG) approach allows for a
non-perturbative solution of gauge theories in terms of its
renormalization group flow for the gauge-fixed effective action. The
underlying gauge symmetry is encoded in modified Ward-Takahashi (mWTI)
or Slavnov-Taylor identities (mSTI) that approach the standard ones in
the limit of a vanishing (infrared) cutoff scale.  A very convenient
representation of the symmetry constraints is given by the Quantum
Master Equation (QME) \cite{Igarashi:2000vf, Igarashi:2001mf,
Igarashi:2007fw, Higashi:2007ax} in the Batalin-Vilkowisky (BV)
antifield formalism \cite{BATALIN198127}.  For fRG reviews on gauge
theories see \cite{Fu:2022gou, Dupuis:2020fhh, Braun:2011pp,
Rosten:2010vm, Igarashi:2009tj, Gies:2006wv, Pawlowski:2005xe}, recent
quantum gravity reviews are collected in \cite{Bambi:2023jiz}, a
discussion of symmetry constraints in quantum gravity can be found in
\cite{Pawlowski:2020qer} and a survey of different fRG approaches in
gauge theories is provided in \cite{Ihssen:2025cff}, for further
developments see \cite{Falls:2025tid}.
 
The gauge-consistent solution of a gauge theory on a renormalized
trajectory (RT) requires approximation schemes for the Wilsonian action
$S$ or the corresponding one-particle irreducible (1PI) effective action
$\Gamma$, whose flow is also a solution of the QME. A further
requirement is the necessary \textit{locality} of the construction. This
constraint is often neglected, for a discussion of local symmetry
operators in the context of chiral symmetry on the lattice see
\cite{Bergner:2012nu}. It has been shown that within given (finite)
truncations there are always non-local completions of the flowing
effective action that also satisfy the mWTIs or mSTIs, for a detailed
discussion see \cite{Dupuis:2020fhh}. In summary, a renormalized
trajectory is a \textit{local} truncation of the effective action whose
flow satisfies the symmetry identities. While perturbation theory
readily offers such a local truncation scheme based on powers of the
gauge coupling, the construction of non-perturbative gauge-consistent
approximation schemes is obstructed by the requirement of solving two
non-linear functional equations simultaneously.

In the present work we aim at the construction of non-perturbative
truncation schemes that lead to renormalized trajectories within a local
implementation of gauge consistency. Within these truncation schemes the
flow equations can be solved numerically on renormalized
trajectories. This programme is based on a chain of works in the past
decade, aiming at perturbative solutions of renormalized trajectories
both in Yang-Mills theory \cite{Igarashi:2019gkm} and in QED
\cite{Igarashi:2021zml}, see also \cite{Igarashi:2016gcf} for the
relation between gauge-consistency of truncations for the Wilsonian
effective action and the 1PI effective action.  In these works the
algebraic structure of the symmetry identities in the BV formalism
played a key role as it linearizes the symmetry constraints in terms of
the antifields.

We extend the study in \cite{Igarashi:2021zml} to non-perturbative
resummed flows. Here, we utilize the crucial observation that the
perturbative construction in \cite{Igarashi:2021zml} readily extends to
non-perturbative flows. This leads us to gauge-consistent fRG-flow in
QED, including a first numerical solution in a qualitative approximation
with four-Fermi terms. Before we set up the construction in detail in
the next Sections, we want to briefly discuss the underlying structure
that allows for this extension: in \cite{Igarashi:2021zml} we have
discussed the one-loop quantum correction of the 1PI effective action,
generated by the expansion of the one-loop formula
\begin{align}
\frac{1}{2}{\rm Str} \log\,\Bigl[\Gamma_{\rm cl}^{(2)} + {\cal R}\Bigr]\,, 
\label{eq:Gammacl-1loop}
\end{align}
where the trace ${\rm Str} $ sums over momenta, Lorentz-, gauge group
and internal indices with a symplectic metric in the fermions. In
\labelcref{eq:Gammacl-1loop}, $\Gamma_{\rm cl}^{(2)}$ is the second
functional derivative of the classical action $\Gamma_{\rm cl}$ with
respect to the fields, and ${\cal R}$ is an infrared regulator term. The
one-loop correlation functions derived from the effective action
$\Gamma$ have been shown to be simultaneous solutions of their flow
equations and the QMEs or modified STIs. \Cref{eq:Gammacl-1loop} can be
readily extended via a RG-improvement to
\begin{align}
\frac{1}{2}{\rm Str} \log\,\Bigl[ (\Gamma_{\rm cl} + \Gamma_{\rm q}^{AA})^{(2)} + {\cal R}\Bigr]\,, 
\label{eq:RG-improved}
\end{align}
where $\Gamma_{q}^{AA}$ are the lowest order correction in the photon
two-point function. \Cref{eq:RG-improved} can be understood as the next
non-perturbative order in a systematic gauge-consistent skeleton
expansion scheme of the flow equation, see
\Cref{app:GenStructureFlow}. The inclusion of such terms accommodates
fully dressed photon propagators, while the fermion propagators remain
classical. Remarkably, the resulting correlation functions solve both,
the QME and the flow equations, as we will explain in this
paper. Although other important quantum corrections are ignored in this
particular truncation, we expect that it will provide a good starting
point for further study, partly because our construction is based on the
algebraic structure mentioned earlier.

Our Ansatz for the 1PI effective action contains the quantum correction
for the photon two-point function.  The correction terms is at
${\cal O}(e^2)$ and satisfies both the flow equation and QME.  This is
consistent with our earlier perturbative result
\cite{Igarashi:2021zml}.

We numerically studied the flows of the couplings using the 1PI
effective action with the above quantum correction, first in the Laudau
gauge.  We find a unique UV fixed point that determines the renormalized
trajectory as well as the phase structure. We have also studied flows
including the flow of the gauge parameter $\xi$ and confirmed the
uniqueness of the UV fixed point obtained in the Landau gauge.  The
linear analysis around the UV fixed point including the $\xi$ flow
equation gives us an extra relevant direction that is responsible for
the $\xi \ne 0$ renormalized trajectory.

QED with four-Fermi couplings has been studied intensively before. These
studies aimed at finding a UV fixed point for the theory, and to
understand the associated chiral symmetry breaking
\cite{Maskawa:1974vs,10.1143/PTP.54.860, Fukuda:1976zb, Miransky:1984ef,
Kondo:1988qd, Bardeen:1985sm, LEUNG1986649, Aoki:1996fh}.  In particular
the authors of \cite{Aoki:1996fh} used the fRG in a different
approximation scheme.  For a comparison to their result, we calculated
the mass anomalous dimension and found a slightly smaller
result. Recently, these studies have been extended to QED with a
Pauli-term, where a UV fixed point has been found \cite{Gies:2020xuh,
Gies:2022aiz}.

Moreover, gauge-consistent approximations within the flow equation
approach has been analyzed extensively, starting with
e.g.~\cite{Bonini:1994kp, Ellwanger:1995qf}. With \Cref{eq:RG-improved}
we can accommodate the momentum dependence of the flows of correlation
functions in a gauge-consistent way. Such momentum-dependent
non-perturbative approximations have also been considered in the past
three decades. They have found use in particular in QCD, where they are
pivotal for quantitative studies of its infrared dynamics. This starts
with \cite{Ellwanger:1995qf, Ellwanger:1996wy, Bergerhoff:1997cv}, for
recent quantitative computations see \cite{Mitter:2014wpa,
Cyrol:2016tym, Cyrol:2017ewj, Corell:2018yil, Fu:2025hcm} and the review
\cite{Dupuis:2020fhh}.

In the next Section, a brief summary of the QME and the flow equation is
given in the 1PI formulation.  In sect. 3 we apply it to QED with chiral
invariant four-Fermi interactions, and show that the correlation
functions computed with generic photon propagators solve the QME.  The
correlation functions are confirmed to obey the flow equation; the
details are explained in \Cref{app:GenStructureFlow}. In
\Cref{sec:dimless_flow_eq} we move on to numerical computations for
$\Lambda$-evolution of anomalous dimensions, gauge and four-Fermi
couplings.  There we use the dimensionless form of the flow equation
with the dressed photon propagator for internal lines and anomalous
dimensions due to the renormalization.  The status of gauge consistency
is explained in \Cref{sec:Gauge_consistency}. A summary is provided in
\Cref{sec:Sum+Dis}.

\section{Flow equation and QME for the 1PI effective action}
\label{sec:Flow+QME}

In this Section we briefly review the set-up within the BV formalism for
the flow equation of the 1PI effective action of a generic gauge theory
in fRG, combined with the quantum master equation.  The latter is
obtained via a Legendre transform from their counterparts for the Wilson
action, as discussed in \cite{Igarashi:2016gcf, Igarashi:2021zml}.

The classical 1PI effective action for a given gauge theory has been
 shown to be the same as the standard one by an analysis of the
 BRST-cohomology.  Its free part takes the form
\begin{align}
\Gamma_{0}[\Phi,\Phi^{*}] = \frac{1}{2} \Phi^{A} \left[\frac{1}{\Delta^{(0)}}\right]_{AB} \Phi^{B} 
+ \Phi_{A}^{*}R^{A}_{~~B} \Phi^{B}\,.
\label{eq:Gamma0_general}
\end{align}
Here $\Phi^{A}$ denote all species of fields, and their kinetic terms
are given by the inverse of the free propagators
$\bigl(\Delta^{(0)}\bigr)^{AB}$. We use DeWitt's condensed notation: the
indices $A,B$ run over space-time or momenta, internal and Lorentz
indices as well as species of fields. The free BRST transformations
$R_{\ B}^A\Phi^B$ for $\Phi^{A}$ are multiplied by their antifields
$\Phi^{*}_{A}$. By construction, they satisfy the relations
\begin{align}
 \left[\frac{1}{\Delta^{(0)}}\right]_{AC} R^C_{\ B} + 
 \left[\frac{1}{\Delta^{(0)}}\right]_{BC}   R^C_{\ A} =0\,.
\label{eq:DeltaRrelation}
\end{align}
The classical BRST-invariant 1PI effective action also includes the classical
interaction $\Gamma_{I, {\rm cl}}$,
\begin{align}
\Gamma_{\rm cl}[\Phi,\Phi^{*}] = \Gamma_{0}[\Phi,\Phi^{*}] 
+ \Gamma_{I,{\rm cl}}[\Phi,\Phi^{*}]\,.
\label{eq:ClassicalAction}
\end{align}
BRST-invariance entails that \labelcref{eq:ClassicalAction} solves the
classical Master Equation (CME), 
\begin{align}
(\Gamma_{\rm cl}, \Gamma_{\rm cl}) =0\,, \qquad \textrm{with}\qquad (X,Y) = \frac{\pa^{r} X}{\pa \Phi^{A}}\frac{\pa^{l} Y}{\pa\Phi^{*}_{A}}
- \frac{\pa^{r} X}{\pa \Phi^{*}_{A}}\frac{\pa^{l} Y}{\pa\Phi^{A}}\,,
\label{eq:CME}
\end{align}
where $\pa^{l(r)}$ denote the left (right) derivative. Once the quantum
part $\Gamma_q$ is included, the 1PI effective action
$\Gamma=\Gamma_{\rm cl}+\Gamma_q$ satisfies the quantum master equation
(QME),
\begin{align} 
\Sigma = \frac{1}{2}(\Gamma, \Gamma) + \Sigma_K=0 \,.
\label{eq:QME_Gamma}
\end{align}
We call the quantity $\Sigma$, defined by the second expression of
\labelcref{eq:QME_Gamma}, the quantum master functional.  $\Sigma_K$ is
written for $\Gamma_I[\Phi, \Phi^*]$, the interaction part of the 1PI
effective action,
\begin{align}
\Sigma_{K} = - {\rm Tr}\left( K \Gamma^{(2)}_{I*}
\frac{1}{1 + \bar\Delta^{(0)} \Gamma^{(2)}_{I}}\right)\,,
\label{eq:Sigma_K}
\end{align}
where 
\begin{align}
\left(\Gamma^{(2)}_I\right)_{AB} = 
\frac{\pa^{l}\pa^{r}}{\pa\Phi^{A}\pa\Phi^{B}}
\,\Gamma_{I}\,,\qquad \qquad 
\left(\Gamma^{(2)}_{I*}\right)^{A}{}_{B} 
= \frac{\pa^{r} \pa^{l}}{\pa \Phi^{*}_{A}
\pa\Phi^{B}}\, \Gamma_{I}\,,
\label{eq:Gamma2_star}
\end{align} 
and the trace $\rm Tr$ sums over includes the multi-indices A, B.  In
the QME in terms of the Wilsonian action, $\Sigma_K$ appears as the
change of the path integral measure under the BRST
transformation. Therefore, we call $\Sigma_K$ the measure term. $K(u)$
is a function of $u = p^2/\Lambda^2$ where $p^2$ is the momentum squared
and $\Lambda$ is a cutoff scale.  As a regulator function, we require
the properties that $K(0)=1$ and $K(u)\to 0$ sufficiently fast as
$u\to\infty$.  Due to the function $K$, $\Sigma_K$ in
\labelcref{eq:Sigma_K} is UV regulated finite. Later we shall observe
cancellations of the antibracket term and the measure term in the QME in
\labelcref{eq:QME_Gamma}.

The flow equation for $\Gamma_{I}$ is given \cite{Wetterich:1992yh}, 
\begin{align}
{\dot \Gamma}_{I} 
= \frac{1}{2}
{\rm Str}\left[\dot{\cal R}\frac{1}{\Gamma^{(2)}+{\cal R} } \right]
= - \frac{1}{2}{\rm Str}\left(\dot{\bar\Delta}^{(0)} 
\frac{1}{\bar\Delta^{(0)}}
\frac{1}{1 + \bar\Delta^{(0)} \Gamma^{(2)}_{I}}\right) \,,
\label{eq:FlowGamma}
\end{align} 
see also \cite{Morris:1993qb,Bonini:1992vh,Ellwanger:1993mw}. The dot
denotes the derivative with respect to $t=\ln \Lambda / \mu$. Here,
$\mu$ is a generic reference scale, typically chosen as the initial
cutoff scale. The function $\cal R$ is
\begin{align} 
{\cal R}_{AB} = \frac{K}{\bar K} \Bigl[\frac{1}{\Delta^{(0)}}\Bigr]_{AB}\,, 
\label{eq:regulator}
\end{align}
with the IR regularization function ${\bar K} = 1-K$. Now we decompose $\Gamma^{(2)}_{I}$ as 
\begin{align}
\Gamma^{(2)}_{I}[\Phi] = \Gamma^{(2)}_{I}[\Phi=0] + \tau[\Phi]\,,
\label{eq:decomposeGamma2}
\end{align}
and define the dressed propagators $\bar \Delta$ including quantum
corrections as
\begin{align}
\frac{1}{\bar\Delta} = \frac{1}{\bar\Delta^{(0)}} + \Gamma^{(2)}_{I}[\Phi=0]\,.
\label{eq:inversepropagator}
\end{align}
The field dependent part $\tau[\Phi]$ gives rise to the interaction
vertices.  Then, the flow equation \labelcref{eq:FlowGamma} can be expressed as
\begin{align}
\dot{\Gamma}_{I} = - \frac{1}{2}{\rm Str}\left[\bar\Delta 
\frac{1}{\bar\Delta^{(0)}}\dot{\bar\Delta}^{(0)} 
\frac{1}{\bar\Delta^{(0)}}
\frac{1}{1 + \bar\Delta \tau[\Phi]}\right] \,
= \frac{1}{2}{\rm Str}\left[\otimes 
\frac{1}{\bar\Delta^{-1}  +  \tau[\Phi]}
\right] \,, 
\label{eq:flow_Gamma_tau}
\end{align}
where the symbol $\otimes$ represents the regulator
\begin{align}
\otimes = {\dot {\cal R}} = - \frac{1}{\bar\Delta^{(0)}}\dot{\bar\Delta}^{(0)} \frac{1}{\bar\Delta^{(0)}} \,.
\label{eq:regulator0}
\end{align}
We shall use the symbol $\otimes$ in \labelcref{eq:regulator0} also in
diagrammatic depictions of the flow. We proceed with a discussion of how
the flow equation \labelcref{eq:flow_Gamma_tau} accumulate higher order
quantum corrections.  Writing the scale dependence explicitly, the flow
equation \labelcref{eq:flow_Gamma_tau} provides an infinitesimal change,
\begin{align}
\Gamma_I(t+\Delta t) \sim \Gamma_I(t) + \Delta t \cdot \frac{1}{2}{\rm Str}\left[\otimes 
\frac{1}{\bar\Delta^{-1}  +  \tau[\Phi]}
\right](t)\,, 
\label{eq:flow_Gamma_int2}
\end{align}
with $t=\log \Lambda/\mu$. If iterating this infinitesimal procedure,
the two-point function $\Gamma^{(2)}_I(t+\Delta t)$ on the right hand
side of \labelcref{eq:flow_Gamma_tau} is the propagator and the vertices
based on \labelcref{eq:decomposeGamma2} and
\labelcref{eq:inversepropagator}.  In this manner, the loop
contributions on the r.h.s. of \labelcref{eq:flow_Gamma_tau} change the
propagator and the vertices in every step of changing the scale.

When we start from an action satisfying the QME and include all the
contributions as explained above, we get a flow that satisfies the QME
at all scales. This is equivalent to the path integral formulation.
Obviously it is extremely difficult to solve the flow equation taking
into account the whole infinite tower of generated quantum
corrections. In \Cref{app:GenStructureFlow} we will discuss the 1PI
action with the RG-improved formula \labelcref{eq:RG-improved} is a
solution to the flow equation with a particular set of interaction
vertices.

The flow equation \labelcref{eq:FlowGamma} or
\labelcref{eq:flow_Gamma_tau} has a closed non-perturbative one-loop
form, which facilitates the discussion of gauge-consistency. We
emphasize, however, that the one-loop structure is deceiving in terms of
perturbation theory: the full field-dependent propagator incorporates
all order of perturbation theory.

\section{1PI effective action and symmetry identities}
\label{sec:1PI+QME}

In the first subsection we construct a 1PI effective action for QED with
chiral invariant four-Fermi interactions including quantum corrections
described by \labelcref{eq:RG-improved}.  The 1PI effective action
differs from the one given in \cite{Igarashi:2021zml} and may be
regarded as an extension of it.

In the succeeding subsections, we discuss the QME for the 1PI effective
action, the vanishing of the quantum master functional denoted as
$\Sigma$ that is a sum of terms classified by powers of the couplings
$e^m G^n$.  We will find that each term $\Sigma_{e^m G^n}$ vanishes by
itself.  Therefore the QME does not produce any relation among the
couplings.  Another interesting result is the fact that the QME holds
for the 1PI effective action with the RG-improved one-loop formula
\labelcref{eq:RG-improved} that produces diagrams with the dressed photon
propagator.

In this Section, all the couplings are regarded as constant.  
Later in \Cref{sec:Gauge_consistency} we will reconsider
the QME in terms of the renormalized 1PI effective action.

\subsection{1PI effective action} 
\label{eq:EffectiveAction}

The free part of 1PI effective action in QED with a massless Dirac fermion 
is given by
\begin{align}\nonumber 
\Gamma_{0} =&\, \int_{x}\Biggl\{
\frac{1}{2}\Bigl[(\pa_{\mu}A_{\nu})^{2}- (\pa\cdot A)^{2}\Bigr]
+ {\bar \Psi} i \Slash{\pa} \Psi \\[1ex]
&\hspace{1.3cm}+ \bigl(A^{*}_{\mu} -i \pa_{\mu} {\bar C}\bigr)\pa_{\mu}C + \frac{1}{2}\xi B^{2}
{+} \bigl({\bar C}^{*} + i \pa\cdot A\bigr)B\Biggr\}\,.
\hspace{1cm}
\label{eq:Gamma0}
\end{align}
It contains kinetic terms for the photon $A_{\mu}$, the massless fermion
$\Psi$ and $\bar\Psi$, and the Faddeev-Popov ghosts $C$ and $\bar C$.
In addition to the fields $\Phi^{A} = \{A_{\mu}, \Psi, \bar\Psi, C, \bar C\}$, 
we also included their antifields $\Phi^{*}_{A} = \{A^{*}_{\mu}, \Psi^{*},
 {\bar \Psi}^{*}, C^{*}, {\bar C}^{*}\}$. 
The auxiliary field $B$ and the gauge parameter $\xi$ are introduced for
a covariant gauge-fixing.  We take the gauge-fixed basis for antifields,
for the notation see \cite{Igarashi:2019gkm}.

The interaction part of the classical action, 
$\Gamma_{I,{\rm cl}}[\Phi,\Phi^{*}]$, has two contributions
\begin{align}
\Gamma_{I,{\rm cl}} = \Gamma_{\rm e} + \Gamma_{\rm G}\,.
\label{eq:GammaI}
\end{align}
The first term $\Gamma_e$ consists of the gauge interaction with the
gauge coupling $e$ and terms generating BRST transformations on the
fermion fields
\begin{align}
 \Gamma_{\rm e} =\int_{x}
\Bigl\{ -e {\bar \Psi} \Slash{A} \Psi -i e \Psi^{*} \Psi C 
+i e {\bar \Psi}^{*} {\bar \Psi} C\Bigr\} \,.
\label{eq:Gamma_e}
\end{align}
We also include chirally  invariant four-Fermi interactions 
\begin{align}\nonumber 
\Gamma_{\rm G} = &\,\int_{x} \Biggl\{\frac{G_{S}}{2}
\Bigl(
\left({\bar\Psi}\Psi\right)\left({\bar\Psi}\Psi\right)
 - \left({\bar\Psi}\gamma_{5}\Psi\right)\left({\bar\Psi}
\gamma_{5}\Psi\right)
\Bigr)\\[1ex]
&\hspace{1cm}
+ \frac{G_{V}}{2}
\Bigl(
\left({\bar\Psi}\gamma_{\mu}\Psi\right)\left({\bar\Psi}\gamma_{\mu}\Psi\right)
 + \left({\bar\Psi}\gamma_{5}\gamma_{\mu}\Psi\right)\left({\bar\Psi}
\gamma_{5}\gamma_{\mu}\Psi\right)
\Bigr)
\Biggr\}\,.
\hspace{1cm}
\label{eq:4fermi}
\end{align}
The classical 1PI effective action $\Gamma_{\rm cl}$,
\begin{align}
\Gamma_{\rm cl} = \Gamma_{0} + \Gamma_{I,{\rm cl}}\,,
\label{eq:Gamma_cl}
\end{align}
is gauge (BRST) invariant or it satisfies the CME \labelcref{eq:CME}. 

In \cite{Igarashi:2021zml}, we computed the quantum part of 1PI
effective action by expanding the one-loop formula ${\rm Str} \log
[\Gamma_{\rm cl}^{(2)} + {\cal R}]$.  The resulting 1PI effective action
is shown to satisfy both, the QME and the flow equation.

In this paper, we consider a non-perturbative extension of
\cite{Igarashi:2021zml} by adding a quantum part of the photon two-point
function to the classical action $\Gamma_{\rm cl}$
\begin{align} 
\Gamma_{\rm q}^{AA} = \frac{1}{2} \int_{x,y} A_{\mu}(x) 
{\cal A}_{\mu\nu}(x-y) A_{\nu}(y) \,,
\label{eq:GammaAA}
\end{align}
and consider the 1PI effective action $\Gamma$ with an RG-improved one-loop formula
\begin{align}\nonumber 
\Gamma =&\, \Gamma_{\rm cl} + \Gamma_q\,,
\\[2ex]
\Gamma_{\rm q} =&\,
\frac{1}{2}{\rm Str} \log  [(\Gamma_{\rm cl} + \Gamma_{\rm q}^{AA})^{(2)} + {\cal R}]\,.
\label{eq:mod.strlog}
\end{align}
\Cref{eq:mod.strlog} produces terms constructed with the vertex functions $\tau[\Phi]$ given
in \Cref{app:vertices}, the IR-regularized free fermion propagator 
\begin{align}
{\bar \Delta}^{(0)}_{\al\hat\beta} = (i \Slash{\pa})_{\al\hat\beta}
\mathbf{\bar \Delta}^{(0)}\,,\qquad \mathbf{\bar \Delta}^{(0)} = {\bar K}/(-\pa^{2})\,,
\label{eq:Dirac_pro}
\end{align}
and the dressed photon propagator as explained in \Cref{eq:inversepropagator}
\begin{align}
\bar\Delta_{\mu\nu} = \left(\bigl({\bar\Delta}^{(0)}\bigr)^{-1} 
+ {\cal A}\right)^{-1}_{\mu\nu}\,.
\label{eq:phton_prop}
\end{align}
In \labelcref{eq:phton_prop}, 
${\bar \Delta}^{(0)}_{\mu\nu}$ is the IR-regularized photon free propagator
\begin{align}
{\bar \Delta}^{(0)}_{\mu\nu} = (P^T_{\mu\nu} + \xi P^L_{\mu\nu})\mathbf{\bar \Delta}^{(0)}\,
\label{eq:IRregPhotonProp}
\end{align}
where $P^{T}_{\mu\nu} = \delta_{\mu\nu} - P^{L}_{\mu\nu},
~P^{L}_{\mu\nu} = \pa_{\mu}\pa_{\nu}/(\pa^{2})$ are projection
operators. By decomposing ${\cal A}_{\mu\nu}$ into the transverse and
longitudinal components as
\begin{align}  
{\cal A}_{\mu\nu} = \left({\cal T} P^{T} + {\cal L}P^{L}\right)_{\mu\nu}\,,
\label{eq:cal_A}
\end{align}
we obtain the dressed photon propagator 
\begin{align}
\bar\Delta_{\mu\nu} = \left(\bigl({\bar\Delta}^{(0)}\bigr)^{-1} 
+ {\cal A}\right)^{-1}_{\mu\nu}
= \bigl( P^{T} T +  P^{L} L \bigr)_{\mu\nu}
\, , 
\label{eq:phton_prop2}
\end{align}
where
\begin{align}
T = \Bigl( - {\bar K}^{-1}\pa^{2} + {\cal T}\Bigr)^{-1}, \qquad 
L = \Bigl( -\xi^{-1} {\bar K}^{-1}\pa^{2} +  {\cal L}\Bigr)^{-1}\,.
\label{eq:TL}
\end{align}
$\Gamma_{\rm q}$ in \labelcref{eq:mod.strlog} produces a series of terms
that may be classified by fields attached and the powers of the
couplings,
\begin{align}
\Gamma_{\rm q}
= \Gamma_{\rm q}^{AA} + \Gamma_{\rm q}^{{\bar \Psi}\Psi} 
+ \Gamma_{\rm q}^{{\bar \Psi} A \Psi} 
+ \Gamma_{\rm q}^{\bar\Psi \Psi\bar\Psi \Psi} + \cdots\,.
\label{eq:mod.strlog2}
\end{align}
Here we only consider terms relevant for later discussions.  In
\labelcref{eq:mod.strlog2}, the superscripts indicate the attached fields to
terms on the r.h.s..  We point out here that terms with internal photon
propagators differ from those given in I.

The quantum part of the photon two-point function is given as
\begin{align}\nonumber 
\Gamma_{\rm q}^{AA} = \Gamma_{e^{2}}^{AA} 
=&\, \frac{1}{2} \bar\Delta^{(0)}_{\al\hat\al}
\tau_{\hat\al\beta}^{(-\Slash{A})}
\bar\Delta^{(0)}_{\beta\hat\beta}\tau_{\hat\beta\al}^{(-\Slash{A})}
\\[1ex]\nonumber 
=&\, \frac{1}{2} e^{2} \int_{x,y}{\rm tr}
\Bigl((i \Slash{\pa})\mathbf{\bar\Delta}^{(0)}(x-y) 
\Slash{A}(y) (i \Slash{\pa})
\mathbf{\bar\Delta}^{(0)}(y-x) \Slash{A}(x) \Bigr)
\\[1ex]
\equiv &\,  \frac{1}{2} e^{2} 
\Bigl[(i \Slash{\pa})\mathbf{\bar\Delta}^{(0)} \Slash{A} (i \Slash{\pa})
\mathbf{\bar\Delta}^{(0)} \Slash{A} \Bigr] \,.
\label{eq:AA}
\end{align}
The last line of \labelcref{eq:AA} is the abbreviated expression of the
second line: the square brackets on the line implies the trace over
spinor indices and the integrations over $x$ and $y$.  The notation
$\Gamma_{e^{2}}^{AA}$ is introduced to show its dependence on the
couplings.  $\Gamma_{e^2}^{AA}$ in \labelcref{eq:AA} is generated by a
fermion loop and it is the same as that in \cite{Igarashi:2021zml}.
\Cref{eq:AA} gives ${\cal A}_{\mu\nu}$ of \labelcref{eq:GammaAA}.

Now we consider $\Gamma_{\rm q}^{{\bar
\Psi}\Psi}$, $\Gamma_{\rm q}^{{\bar \Psi} A \Psi}$ and $\Gamma_{\rm
q}^{\bar\Psi \Psi\bar\Psi \Psi}$ for which the dressed photon propagator
\labelcref{eq:phton_prop2} 
may appear in internal lines.  
\begin{figure}
   \centering
   \includegraphics[width=0.55\columnwidth]{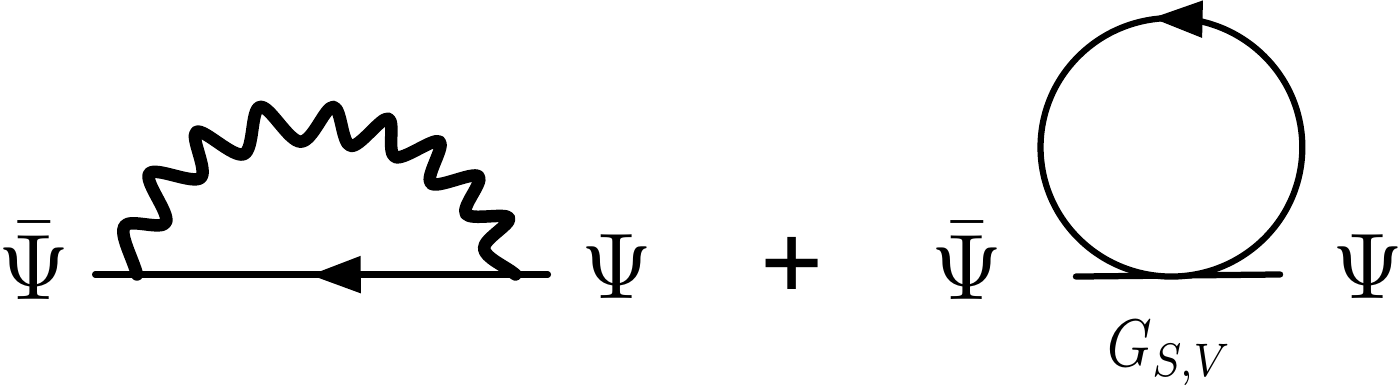}
\caption{Contributions to $\Gamma^{\bar\Psi \Psi}_{\rm q}$.
The thick wavy internal line represents the dressed photon propagator in
\labelcref{eq:phton_prop2}.  The second term vanishes as explained in I.}
\label{fig:2pt_fermion}
\end{figure}
The fermion two-point function consists of two terms as depicted in
\Cref{fig:2pt_fermion}.  Though these are the same diagrams as those in
\cite{Igarashi:2021zml}, we now use the dressed photon propagator
\labelcref{eq:phton_prop2} for an internal photon line shown with the
thick wavy line in \Cref{fig:2pt_fermion}.  As explained in
\cite{Igarashi:2021zml}, the second diagram vanishes.  Therefore the
correction to the fermion two-point function becomes
\begin{align}
\Gamma_{\rm q}^{\bar\Psi\Psi} 
= \Gamma_{ e^{2}}^{\bar\Psi\Psi}=
\bar\Delta^{(0)}_{\al\hat\al} \tau_{\hat\al\mu}^{(-\gamma\Psi)} 
\bar\Delta_{\mu\nu} 
\tau_{\nu\al}^{(- \bar\Psi\gamma)}
= 
- e^{2} \Bigl[\bar\Delta_{\mu\nu}\bar\Psi\gamma_{\nu} 
(i \Slash{\pa}) \mathbf{\bar\Delta}^{(0)} \gamma_{\mu} \Psi \Bigr]\,.
\label{eq:barPsi_Psi}
\end{align}
\begin{figure}[b]
  \begin{minipage}[b]{0.45\linewidth}
    \centering
    \includegraphics[keepaspectratio, scale=0.32]{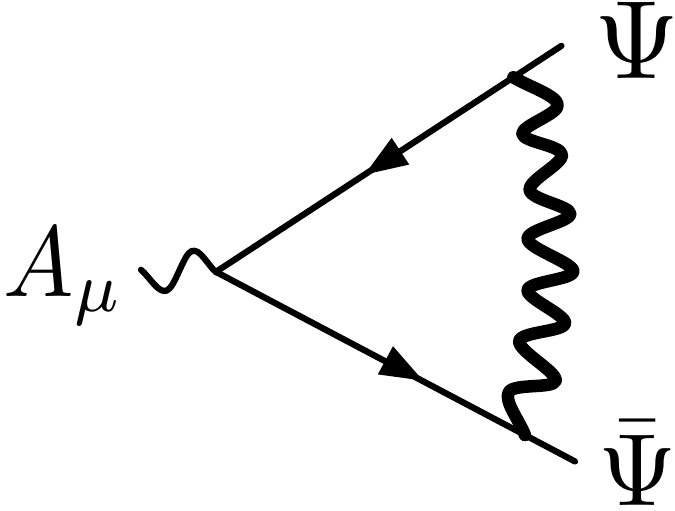}
    \subcaption{$e^3$ contribution, $\Gamma_{e^{3}}^{\bar\Psi A \Psi}$}
  \end{minipage}
  \begin{minipage}[b]{0.45\linewidth}
    \centering
    \includegraphics[keepaspectratio, scale=0.32]{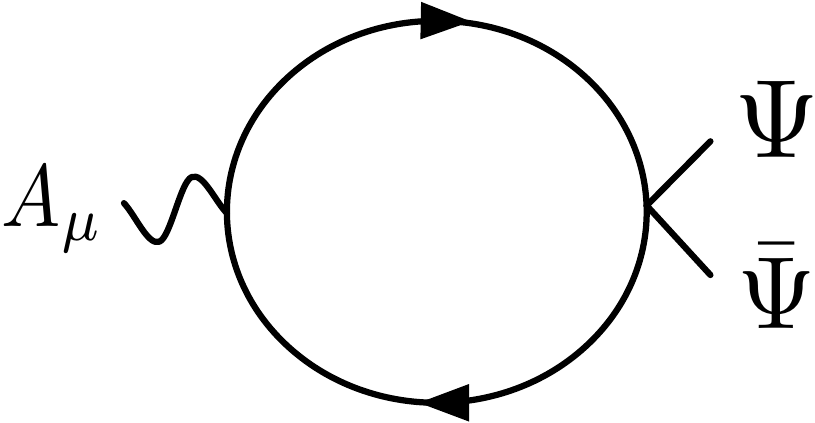}
    \subcaption{$eG$ contribution, $\Gamma_{eG}^{\bar\Psi A \Psi}$}
\label{eG diagram}
  \end{minipage}
  \caption{All the internal propagators are IR regulated.}
\label{fig:q-corrections to 3pt}
\end{figure}
There are two types of quantum corrections to ${\bar\Psi} \Slash{A}
\Psi$ vertex as shown in \Cref{fig:q-corrections to 3pt}, 
\begin{align}
\Gamma_{\rm q}^{\bar\Psi A \Psi} = 
\Gamma_{e^{3}}^{\bar\Psi A \Psi} 
+ \Gamma_{e{\rm G}}^{\bar\Psi A \Psi}\,.
\label{eq:barPsi_A_Psi}
\end{align}
The first term $\Gamma_{e^{3}}^{\bar\Psi A \Psi} $ is generated solely
with the gauge interaction and
\begin{align}\nonumber 
\Gamma_{e^{3}}^{{\bar\Psi} A \Psi}
=&\,  \frac{1}{3}\Biggl[- \bar\Delta^{(0)}_{\al\hat\al} \bigl(
\tau_{\hat\al\mu}^{(-\gamma\Psi)} \bar\Delta_{\mu\nu} 
\tau_{\nu\beta}^{(- \bar\Psi\gamma)}\bar\Delta^{(0)}_{\beta\hat\beta} 
\tau_{\hat\beta\al}^{(-\Slash{A})} 
- \tau_{\hat\al\beta}^{(-\Slash{A})} \bar\Delta^{(0)}_{\beta\hat\beta}
\tau_{\hat\beta\mu}^{(-\gamma\Psi)}\bar\Delta_{\mu\nu} 
\tau_{\nu\al}^{(- \bar\Psi\gamma)} \bigr)
\\[1ex]\nonumber 
&\hspace{.5cm} + \bar\Delta_{\mu\nu} \tau_{\nu\al}^{(- \bar\Psi\gamma)}
\bar\Delta^{(0)}_{\al\hat\al}\tau_{\hat\al\beta}^{(-\Slash{A})} 
\bar\Delta^{(0)}_{\beta\hat\beta} \tau_{\hat\beta\mu}^{(-\gamma\Psi)}\Biggr]\\[1ex]
=&   
-e^{3}\Bigl[ \bar\Delta_{\mu\nu} \bar\Psi \gamma_{\nu} 
(i\Slash{\pa})\mathbf{\bar\Delta}^{(0)}\Slash{A} (i\Slash{\pa})
\mathbf{\bar\Delta}^{(0)}\gamma_{\mu} \Psi\Bigr]\,.
\label{eq:Gamma3_e3}
\end{align}
The second term $\Gamma_{eG}^{\bar\Psi A \Psi}$ in
\labelcref{eq:barPsi_A_Psi} with four-Fermi interactions is given by
\begin{align}\nonumber 
\Gamma_{e {\rm G}}^{\bar\Psi A \Psi}
=&\, \bar\Delta^{(0)}_{\al\hat\al}
\tau^{(-\Slash{A})}_{\hat\al\beta}\bar\Delta^{(0)}_{\beta\hat\beta}
\tau^{\bar\Psi\Psi}_{\hat\beta\al} 
\\[1ex]\nonumber 
&\hspace{-1cm}= 2 e\, {G_{S}}\Bigl[\bar\Psi (i \Slash{\pa})
{\mathbf {\bar\Delta}}^{(0)} \Slash{A} (i \Slash{\pa})
{\mathbf {\bar\Delta}}^{(0)} 
\Psi\Bigr]
+ 2e {G_{V}} \Bigl[\bar\Psi \gamma_{\mu} 
(i \Slash{\pa}){\mathbf {\bar\Delta}}^{(0)} \Slash{A} 
(i \Slash{\pa}){\mathbf {\bar\Delta}}^{(0)}
\gamma_{\mu} \Psi \Bigr]\, \\[1ex]
&\hspace{-.6cm} -4 e {G_{V}}
\Bigl[i\pa_{\rho} \bar{\mathbf \Delta}^{(0)} A_{\rho} i \pa_{\mu}\bar{\mathbf \Delta}^{(0)}
- i\pa_{\rho} \bar{\mathbf \Delta}^{(0)} A_{\mu} i \pa_{\rho}\bar{\mathbf \Delta}^{(0)}
+ i\pa_{\mu} \bar{\mathbf \Delta}^{(0)} A_{\rho} i \pa_{\rho}\bar{\mathbf \Delta}^{(0)}
\Bigr]\Bigl(\bar\Psi \gamma_{\mu} \Psi\Bigr)\,.
\label{eq:Gamma3_eG}
\end{align}

   \begin{figure}[t]
        \centering
    \begin{tabular}{cc}
      \begin{minipage}[t]{0.4\hsize}
        \centering
        \includegraphics[keepaspectratio, scale=0.3]{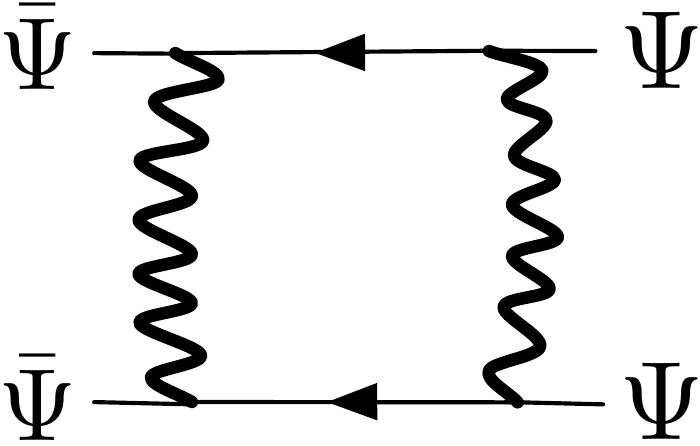}
        \subcaption{$\Gamma_{e^{4}}^{\bar\Psi \Psi\bar\Psi \Psi}$}
       \label{four-Fermi_e4}
      \end{minipage} &
       \begin{minipage}[t]{0.4\hsize}
         \centering
         \includegraphics[keepaspectratio, scale=0.3]{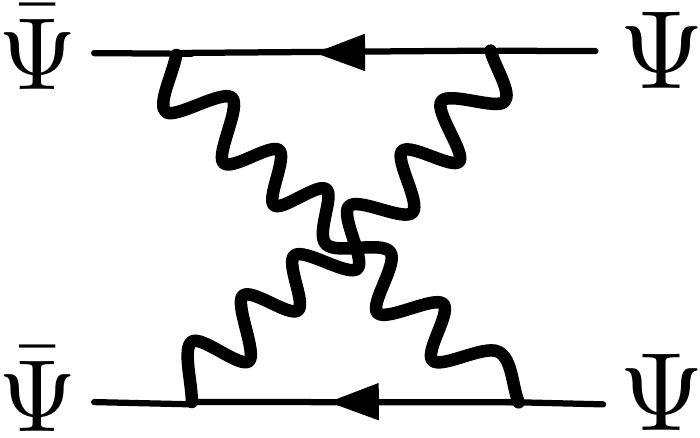}
         \subcaption{$\Gamma_{e^{4}}^{\bar\Psi \Psi\bar\Psi \Psi}$, non-planar contributions}
         \label{four-Fermi_e4-2}
       \end{minipage} \\ 
      \begin{minipage}[t]{0.4\hsize}
        \centering
        \includegraphics[keepaspectratio, scale=0.3]{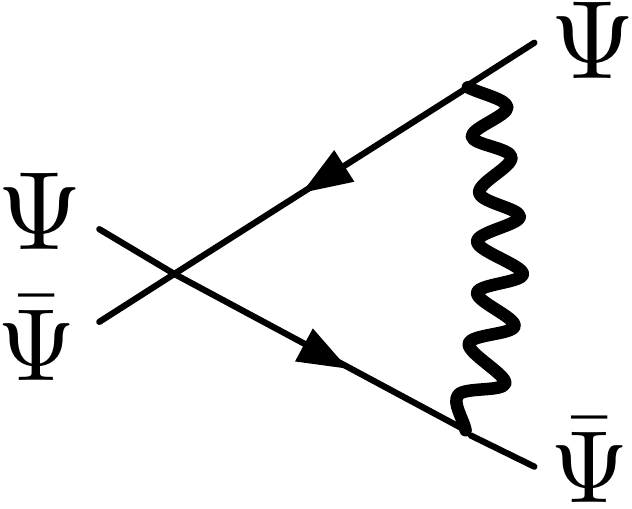}
        \subcaption{$\Gamma_{e^{2} {\rm G}}^{\bar\Psi \Psi\bar\Psi \Psi}$}
        \label{e2G}
      \end{minipage} &
      \begin{minipage}[t]{0.4\hsize}
        \centering
        \includegraphics[keepaspectratio, scale=0.3]{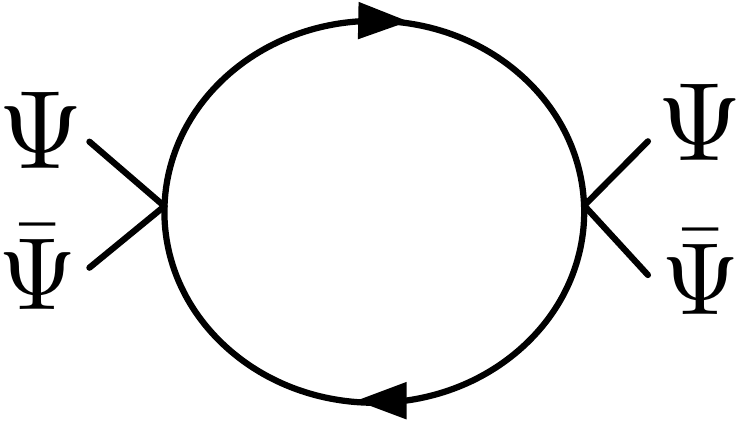}
        \subcaption{$\Gamma_{{\rm G}^{2}}^{\bar\Psi \Psi\bar\Psi \Psi} $.}
        \label{GG}
      \end{minipage} 
    \end{tabular}
        \centering
   \caption{RG-one-loop diagrams for $\Gamma_{\rm q}^{\bar\Psi \Psi\bar\Psi \Psi}$.  There are variations
in connecting fermion lines to four-Fermi vertices other than
Figs. \ref{e2G} and \ref{GG}.}
\label{fig:Gamma-4fermi-q}
\end{figure}
We emphasize that $\Gamma_{e^{2}}^{\bar\Psi\Psi}$ and
$\Gamma_{e^{3}}^{{\bar\Psi} A \Psi}$ have the dressed photon propagators
\labelcref{eq:phton_prop} while the loops for $\Gamma_{e^{2}}^{AA}$ and
$\Gamma_{e {\rm G}}^{\bar\Psi A \Psi}$ contain only the IR regulated
free fermion propagators \labelcref{eq:Dirac_pro} and, therefore, are the same
as those given in I.

The quantum corrections to four-Fermi interactions consist of the
contributions proportional to $e^4$, $e^{2}{\rm G}$ and ${\rm G}^{2}$:
$\Gamma_{\rm q}^{\bar\Psi \Psi\bar\Psi \Psi}=
\Gamma_{e^{4}}^{\bar\Psi \Psi\bar\Psi \Psi}
+ \Gamma_{e^{2} {\rm G}}^{\bar\Psi \Psi\bar\Psi \Psi}
+ \Gamma_{{\rm G}^{2}}^{\bar\Psi \Psi\bar\Psi \Psi}$.
Since they are not necessary in our discussions for QME, we only show
diagrams in \Cref{fig:Gamma-4fermi-q}.  These diagrams will be
helpful to understand the discussions in \Cref{Flows of
couplings}.  There are variations in connecting fermion
lines to four-Fermi vertices other than Figs. \ref{e2G} and \ref{GG}.

\subsection{QME for the $AC$-sector}
\label{sec:QME-AC}

The Quantum Master Functional $\Sigma$ in \Cref{eq:QME_Gamma} can
be decomposed into a sum of products of fields. In QED, each product
consists of the ghost $C$ multiplied by a polynomial of the photon
fields $A_{\mu}$ and/or the fermion fields $\Psi$ and $\bar\Psi$.  Here we
consider two types of products, $A C$ and ${\bar \Psi} \Psi C$, denoted
as $\Sigma^{AC}$ and $\Sigma^{{\bar \Psi} \Psi C}$.

Let us first consider $\Sigma^{A C}$. It is given by 
\begin{align}
\Sigma^{A C} = \Sigma^{A C}_{e^{2}} =(\Gamma_{0},~ \Gamma_{\rm q}^{AA}) 
+ \Sigma_K^{AC}\,.
\label{eq:Sigma_2}
\end{align}
Since all propagators that appear on r.h.s. are fermionic, $\Sigma^{A C}_{e^{2}}$ is 
the same as the one computed in \cite{Igarashi:2021zml} that was shown to vanish $\Sigma^{A C}_{e^{2}}=0$.

The condition $\Sigma^{A C}_{e^{2}}=0$ determines the
longitudinal component $\cal L$ in \labelcref{eq:cal_A} from its measure term
$\Sigma_K^{AC}$.  $\cal L$ calculated from the one-loop formula
\labelcref{eq:mod.strlog} gives the same function as the one from $\Sigma^{A
C}_{e^{2}}=0$.  For later use, we give the expression for $\cal L$
given in \cite{Igarashi:2016gcf} and I:
\begin{align}
{\cal L}(p) = -8 e^2 \int_q K(p+q){\bar K}(q) \frac{(p \cdot q)}{p^2 q^2} \,,\qquad \quad
\int_q = \int \frac{d^4 q}{(2 \pi)^4}\,.
\label{eq:calT+L_QME}
\end{align}
Having the regulator functions $K$ and ${\bar K}$, the integral in
\labelcref{eq:calT+L_QME} gives a finite result.  In the limit
$\Lambda\to 0$, the function $\cal L$ vanishes as expected.

\subsection{QME for the ${\bar \Psi} \Psi C$ sector}
\label{sec:barPsi_Psi_C}

The quantum master functional for the ${\bar\Psi\Psi C}$ sector consists
of two parts distinguished by powers of couplings 
\begin{align}
\Sigma^{\bar\Psi\Psi C} = \Sigma^{\bar\Psi\Psi C}_{e^{3}} + 
\Sigma^{\bar\Psi\Psi C}_{e {\rm G}}\,.
\label{eq:Sigma_3}
\end{align}
It follows from (\ref{eq:QME_Gamma}) that 
\begin{align}\nonumber 
\Sigma^{\bar\Psi\Psi C}_{e^{3}} =&\, 
(\Gamma_{e}, \Gamma_{e^{2}}^{\bar\Psi\Psi}) + 
(\Gamma_{0}, \Gamma_{e^{3}}^{\bar\Psi A  \Psi})
+  \Sigma^{\bar\Psi\Psi C}_{K~e^{3}}\,,\\[2ex]
\Sigma^{\bar\Psi\Psi C}_{eG} =&\,
(\Gamma_{0}, \Gamma_{eG}^{\bar\Psi A  \Psi})
+  \Sigma^{\bar\Psi\Psi C}_{K~eG}\,.
\label{eq:Sigma_eG+Sigma_3e}
\end{align}
$\Sigma^{\bar\Psi\Psi C}_{e{\rm G}}$ in \labelcref{eq:Sigma_eG+Sigma_3e} involves only
fermion loops and we already showed its vanishing in \cite{Igarashi:2021zml}.

The antibracket contributions in the first equation in
\labelcref{eq:Sigma_eG+Sigma_3e} are given as
\begin{align}\nonumber 
(\Gamma_{e}, \Gamma_{e^{2}}^{\bar\Psi\Psi}) =&\, 
e^{3} \Bigl[\bar\Psi\gamma_{\nu}  \Slash{\pa}
\mathbf{\bar\Delta}^{(0)} C\gamma_{\mu} \Psi \bar\Delta_{\mu\nu}\Bigr]
- e^{3}\Bigl[ \bar\Psi  \gamma_{\nu}C  \Slash{\pa}
\mathbf{\bar\Delta}^{(0)}
\gamma_{\mu} \Psi \bar\Delta_{\mu\nu}\Bigr]\,,
\\[2ex]\nonumber 
(\Gamma_{0},\Gamma_{e^{3}}^{\bar\Psi A  \Psi}) =&\, 
- e^{3} \Bigl[
\bar\Psi \gamma_{\nu} \Slash{\pa}\mathbf{\bar\Delta}^{(0)}
\Slash{\pa} C \Slash{\pa} \mathbf{\bar\Delta}^{(0)}
\gamma_{\mu} \Psi\bar\Delta_{\mu\nu}\Bigr]
\\[1ex]
 =&\, e^{3} 
\Bigl[\bar\Psi \gamma_{\nu} (1-K) C \Slash{\pa}\mathbf{\bar\Delta}^{(0)}
\gamma_{\mu} \Psi\bar\Delta_{\mu\nu}\Bigr] - e^{3} \Bigl[
\bar\Psi \gamma_{\nu} \Slash{\pa} \mathbf{\bar\Delta}^{(0)}
C (1-K)
\gamma_{\mu} \Psi \bar\Delta_{\mu\nu} \Bigr]\,,
\label{eq:S3s0}
\end{align}
by using integration by parts and $\pa^{2} \mathbf
{\bar\Delta}^{(0)} = - {\bar K}$. 
The measure part reads 
\begin{align}\nonumber 
\Sigma^{\bar\Psi\Psi C}_{K~e^{3}}
=&\, -   \bigl[K \tau^{C}_{*\al\beta} \bar\Delta^{(0)}_{\beta\hat\al} 
\tau_{\hat\al\mu}^{(-\gamma\Psi)}\bar\Delta_{\mu\nu} 
\tau_{\nu\al}^{(-\bar\Psi\gamma)}  + K 
\tau^{C}_{*\hat\al\hat\beta} \bar\Delta^{(0)}_{\hat\beta\al} 
\tau_{\al\mu}^{(\bar\Psi\gamma)}\bar\Delta_{\mu\nu} 
\tau_{\nu\hat\al}^{(\gamma\Psi)}\bigr] 
\\[1ex]
=&\,
e^{3} \Bigl[\bar\Psi\gamma_{\nu} KC \Slash{\pa}\mathbf{\bar\Delta}^{(0)}
\gamma_{\mu} \Psi \bar\Delta_{\mu\nu}\Bigr] -  e^{3} \Bigl[\bar\Psi\gamma_{\mu} 
\Slash{\pa} \mathbf{\bar\Delta}^{(0)} C K  \gamma_{\nu} 
\Psi\bar\Delta_{\mu\nu}\Bigr]\,. 
\label{eq:Sigma3K}
\end{align}
By summing up \labelcref{eq:S3s0} and \labelcref{eq:Sigma3K}, we confirm that 
\begin{align}
\Sigma^{\bar\Psi\Psi C}_{e^{3}} = 
(\Gamma_{e},\Gamma_{e^{2}}^{\bar\Psi\Psi})
+ (\Gamma_{0},\Gamma_{e^{3}}^{\bar\Psi A  \Psi})
+ \Sigma^{\bar\Psi\Psi C}_{K~e^{3}} =0\,.
\label{Sigma3e}
\end{align}
Therefore, $\Sigma^{\bar\Psi\Psi C} = \Sigma^{\bar\Psi\Psi C}_{e^{3}} 
+ \Sigma^{\bar\Psi\Psi C}_{e{\rm G}} =0$.  

In conclusion, we have shown $\Sigma^{AC} = \Sigma^{\bar\Psi\Psi C} =0$.
Let us point out here that the photon propagator $\bar\Delta_{\mu\nu}$ behaves as
a spectator in the proof of $\Sigma^{\bar\Psi\Psi C} =0$: a particular
form of the photon propagator $\bar\Delta_{\mu\nu}$ is not required to
show the condition.  

Note that the QME for the 1PI effective action (\ref{eq:mod.strlog})
does not produce any relations among the couplings. The authors of
\cite{PhysRevLett.93.110405} and \cite{Gies:2003dp} obtained a
constraint between the gauge and four-Fermi couplings out of the
modified Ward-Takahashi identity and discussed its physical
implications.  In our earlier publication \cite{Igarashi:2016gcf}, we
also had a relation among couplings when we keep generic form factors
for various vertices.  This implies that constraints among couplings out
of the QME depend on an Ansatz for 1PI effective action.  We may
understand that the 1PI effective actions in the present work and our
perturbative studies \cite{Igarashi:2019gkm, Igarashi:2021zml} have
particular form factors so that each of the terms, classified by the
powers $e^n G_{S,V}^{m}$, vanishes by itself. This is an intriguing
possibility and a comprehensive discussion will be provided elsewhere.

\section{Numerical study}
\label{sec:dimless_flow_eq}

Here we introduce our Ansatz including the quantum correction for the
photon two-point function, which give rise the momentum dependent form
factor. This is preceded by a derivation of the dimensionless form of
the flow equation.  The numerical results are explained and compared
with earlier studies.

\subsection{Z factors and inclusion of anomalous dimensions}
\label{sec:Z+AmonalousDim}

Let us introduce anomalous dimensions into the flow equations, rescaling
the fields as $\Phi^{A} = Z_{A}^{1/2} \Phi_{(r)}^{A}$ with the wave function
renormalization $Z_{A}$. The regulator terms (\ref{eq:regulator})
are replaced by $Z_{A}{\cal R}_{AB}$ on which $\Lambda \pa_{\Lambda}
=\pa_{t}$ derivative on the r.h.s. of the flow equations \Cref{eq:FlowGamma}
acts as 
\begin{align}
\pa_{t} \bigl(Z_{A} {\cal R}_{AB} \bigr) = Z_{A}\bigl[ {\dot{\cal R}}_{AB} 
- \eta_{A} {\cal R}_{AB}\bigr]\,,
\label{eq:pa_tCalR}
\end{align}
where 
\begin{align}
\eta_{A} = - \pa_{t} \ln Z_{A}\,,
\label{eq:anom_orig}
\end{align}
denote anomalous dimensions for $\Phi^{A}$.  For our correlation functions, 
$\Gamma_{\rm q}^{AA}, \Gamma_{\rm q}^{\bar\Psi\Psi},  
\Gamma_{\rm q}^{\bar\Psi A \Psi}$ and 
$\Gamma_{\rm q}^{\bar\Psi\Psi\bar\Psi\Psi}$, 
$\pa_{t}$ acts on the free 
propagators. Since 
\begin{align}\nonumber 
 \dot{\cal R} - \eta {\cal R}= &\, 
(\pa_{t}-\eta) \frac{K}{\bar K}\bigl(\Delta^{(0)}\bigr)^{-1} 
= \pa_{t}\bigl(\bar\Delta^{(0)}\bigr)^{-1} - \eta K 
\bigl(\bar\Delta^{(0)}\bigr)^{-1}
\\[1ex]
 =&\,  - \bigl(\bar\Delta^{(0)}\bigr)^{-1}\Bigl[\bigl(\pa_{t} + \eta K\bigr)
\bar\Delta^{(0)}\Bigr]\bigl(\bar\Delta^{(0)}\bigr)^{-1}\,,
\label{eq:eta_replacement}
\end{align}
the anomalous dimensions can be included by the replacement $\pa_{t}
\bar\Delta^{(0)} \to \bigl(\pa_{t} + \eta K\bigr)
\bar\Delta^{(0)}$. Note that the factor $K$ multiplied by $\eta$ makes
the flow equations UV finite. Using this prescription, we show that our
correlation functions fulfill the flow equations with anomalous
dimensions. In our case, the corresponding $Z$ factors are defined as
\begin{align}\nonumber 
 A_{(r)\mu} =&\, Z_{3}^{1/2} A_{\mu}\,, 
&\quad 
A^{*}_{(r)\mu} =&\, Z_{3}^{-1/2} A^{*}_{\mu},
&\quad B_{(r)} =&\, Z_{3}^{-1/2} B\,, 
\\[1ex]\nonumber 
C_{(r)} =&\, Z_{3}^{1/2} C\,, 
& {\bar C}_{(r)} =&\, Z_{3}^{-1/2} {\bar C} \,, 
& {\bar C}^{*}_{(r)} =&\, Z_{3}^{1/2}{\bar C}^{*}\,, 
\\[1ex]
\Psi_{(r)} =&\, Z_{2}^{1/2} \Psi\,,
& \Psi^{*}_{(r)} =&\, Z_{2}^{-1/2} \Psi^{*}\,, 
&{\bar\Psi}_{(r)} =&\, Z_{2}^{1/2}\bar\Psi, 
& {\bar\Psi}^{*}_{(r)} =&\, Z_{2}^{-1/2} \Psi^{*}
\,.
\label{Z_factor}
\end{align}
Note that the transformations $\Phi^{A}_{(r)}= Z_{A}^{1/2}\Phi^{A}$ and 
$\Phi^{*}_{(r)A} = Z_{A}^{-1/2}\Phi^{*}_{A}$  in 
(\ref{Z_factor}) are canonical transformations in the space of fields and 
antifields.
For the couplings and the gauge parameter, we set 
\begin{align}
e_{(r)} = e(t)=Z_{e} e, 
\qquad G_{(r)S,V} = G_{S,V}(t) =Z_{S,V} \cdot \Lambda^2 G_{S,V}, 
\qquad \xi_{(r)} = \xi(t) =Z_{3}\xi \,. 
\label{eq:couplings}
\end{align}
Here we introduced the notation $e(t)$, $G_{S,V}(t)$ and $\xi(t)$ for
the corresponding renormalized quantities for later convenience. After
making all the quantities dimensionless, we find the flow equation in
four-dimensional momentum space as
\begin{align}\nonumber 
 \pa_{t} \Gamma [\hat\Phi_{(r)}]\mid_{\hat\Phi_{(r)}}
=&\, \frac{1}{2}
(-)^{\ep_{A}} \int_{\hat{p}}
\bigl[\pa_{t} \hat{\cal R}- \eta~ \hat{\cal R} \bigr]_{AB}
\left[ 
\left(\frac{\pa^{l} \pa^{r} \Gamma}{\pa \hat\Phi_{(r)}\pa \hat\Phi_{(r)} }
+ \hat{\cal R} \right)^{-1}
\right]^{BA}(\hat{p}, \hat{p}) - 4~ {\Gamma}[{\hat\Phi_{(r)}}]\\[1ex]
&\hspace{.3cm}+ \left(d_{A} + \eta_{A}/2\right)\int_{\hat{p}}\hat\Phi_{(r)}^{A}(\hat{p}) 
\frac{\pa^{l}\Gamma[{\hat\Phi_{(r)}}]}{\pa \hat\Phi_{(r)}^{{A}}(\hat{p})} 
 +  \int_{\hat{p}}
\hat\Phi_{(r)}^{A}(\hat{p})~ \hat{p} \cdot 
\frac{\pa^{\prime}}{\pa \hat{p}} 
\left(\frac{\pa^{l} \Gamma[{\hat\Phi_{(r)}}]}{\pa \hat\Phi_{(r)}^{{A}}(\hat{p})}\right)\,,
\label{eq:wetterich}
\end{align}
where $\hat{p} = p/\Lambda$, $\hat{\cal R}_{AB} =
\Lambda^{d_{A}+d_{B}-d} {\cal R}_{AB}$, and $\hat\Phi_{(r)}^{A}(\hat{p})
= \Lambda^{d-d_{A}}\Phi^{A}_{(r)}(p)$ for the renormalized fields
$\Phi^{A}_{(r)}(p)$ with canonical dimensions $d_{A}$.  The
$t$-derivative on the l.h.s. is now taken for fixed $\hat\Phi_{(r)}^{A}$.
The prime of $\pa^{\prime}/\pa \hat{p}$ on the r.h.s. implies that the
derivative does not act on the delta function for momentum conservation.
$\pa_{t} \hat{\cal R}- \eta~ \hat{\cal R}$ is the dimensionless form of
the quantity defined in \labelcref{eq:regulator0} and the same symbol $\otimes$
will be used to represent it.

We used the hatted notations to denote dimensionless quantities
in \labelcref{eq:wetterich}.  From now on we drop the hats for simplicity. 
In our 1PI effective
action in \Cref{sec:1PI+QME}, the antifield appears only to define
the BRST transformations and no 1PI diagrams are present to change the
terms.  The scale changes of the terms with antifields are due to the Z
factors described above and we can easily recover those scale changes if necessary.
Based on these observations, we ignore the terms with antifields in this Section.

Note that the quantum corrections collected by using
\labelcref{eq:wetterich} differ from \labelcref{eq:RG-improved}.  See
\Cref{app:RG-Improvement} and \Cref{app:dimless flow} for details.

\subsection{Ansatz for the dimensionless 1PI effective action}
\label{sec:AnsatzG}

From \labelcref{eq:Gamma0} we eliminate all antifields $\Phi^{*}_{A}=0$ and $B$
field to obtain the free action 
\begin{align}
\Gamma_{0} =  \int_{p}\biggl[\frac{1}{2} 
A_{\mu}(-p)\Bigl[\delta_{\mu\nu}p^{2} 
+ (\xi(t)^{-1} -1) p_{\mu}p_{\nu}\Bigr]A_{\nu}(p)
+ {\bar \Psi}(-p) \Slash{p} \Psi(-p)
\biggr]\,.
\label{free_dimless_action}
\end{align}
The classical interaction part is given by
\begin{align}\nonumber 
 \Gamma_{I,{\rm cl}} =&\, -e(t) \int\limits_{p,q} 
{\bar \Psi}(-p) \Slash{A}(p-q) \Psi(q) 
+ \frac{1}{2}
\int\limits_{p_{1},\cdots,p_{4}}(2\pi)^{4}
\delta^{4}(p_{1}+p_{2}+p_{3}+p_{4})   
\\[1ex]\nonumber 
&\hspace{.3cm}  \times\Biggl[ 
G_{S}(t)
\Bigl\{\left(
{\bar\Psi}(p_{1})\Psi(p_{2})\right)\left({\bar\Psi}(p_{3})\Psi(p_{4})\right)
 - \left({\bar\Psi}(p_{1})\gamma_{5}\Psi(p_{2})\right)\left({\bar\Psi}(p_{3})
\gamma_{5}\Psi(p_{4})\right)\Bigr\}
\\[1ex]
&\hspace{-1.4cm}  + G_{V}(t)\Bigl\{\left(
{\bar\Psi}(p_{1})\gamma_{\mu}\Psi(p_{2})\right)
\left({\bar\Psi}(p_{3})\gamma_{\mu}
\Psi(p_{4})\right)
+ \left({\bar\Psi}(p_{1})\gamma_{5}\gamma_{\mu}\Psi(p_{2})\right)
\left({\bar\Psi}(p_{3})
\gamma_{5}\gamma_{\mu}\Psi(p_{4})\right)\Bigr\}
\Biggr]\,.
\label{classical_int_action}
\end{align}
If we take the classical 1PI effective action 
$\Gamma_{\rm cl}=\Gamma_{0} + \Gamma_{I,{\rm cl}}$, 
$[\Gamma_{\rm cl}^{(2)} + {\cal R}]^{-1}$ in \labelcref{eq:wetterich}
can be expanded in terms of the IR-regularized free propagators, 
${\bar \Delta}^{(0)}_{\al\hat\beta}$ and ${\bar \Delta}^{(0)}_{\mu\nu}$, and the 
vertices generated from $\Gamma_{I,{\rm cl}}$. 
It leads to a truncation of the flow equations 
at the lowest order in couplings. 

\begin{figure}
 \begin{center}
  \includegraphics[width=4.5cm]{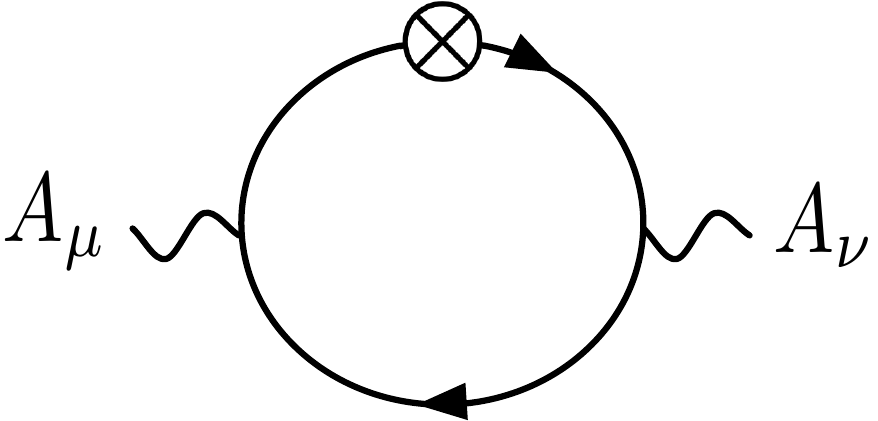}
 \end{center}
 \caption{The above one-loop diagram contributes to the photon two-point function.  The symbol $\otimes$ represents the regulator $\partial_t {\cal R}-\eta {\cal R}$ in the flow equation \labelcref{eq:wetterich}.}  
\label{fig:2pt photon}
\end{figure}
As discussed in \Cref{sec:1PI+QME}, we proceed beyond the
lowest order computation by including the quantum correction to the
photon two-point function:
\begin{align}\nonumber 
 \Gamma_{e^2}^{AA} =&\, \frac{1}{2}\int_{p} A_{\mu}(-p) {\cal A}_{\mu\nu}(p)
A_{\nu}(p)\,,
\\[1ex]
{\cal A}_{\mu\nu}(p,t) = &\, P^{T}_{\mu\nu} {\cal T}(p^{2},t) + P^{L}_{\mu\nu}
{\cal L}(p^{2},t)\,.
\label{eq:cal_A_again}
\end{align}
The two functions ${\cal T}(p^2, t)$ and ${\cal L}(p^2, t)$ are computed from their flows. 

Finally, our Ansatz for the 1PI effective action is given by  
\begin{align}
\Gamma = \Gamma_{\rm cl} + \Gamma_{e^2}^{AA}\,.
 \label{eq:1PI_for_flow}
\end{align}
With the inclusion of ${\cal A}_{\mu\nu}$ the IR-regularized free 
propagators ${\bar \Delta}^{(0)}_{\mu\nu}$ are replaced by the dressed one, ${\bar
\Delta}_{\mu\nu} = \left(\bigl({\bar\Delta}^{(0)}\bigr)^{-1} + {\cal
A}\right)^{-1}_{\mu\nu}$.   

With the Ansatz \labelcref{eq:1PI_for_flow}, the r.h.s. of the flow
equation \labelcref{eq:wetterich} is constructed with the dressed photon
propagator, the free fermion propagator and the classical interaction
vertices that give rise the same quantum corrections discussed in
\Cref{sec:1PI+QME} with appropriately inserted regulator terms.  Though
we do not have momentum-dependent form factors for the fermion two-point
function and interaction vertices, this does not imply that we ignore
all the quantum corrections to them. In writing down the flow equations
for the couplings, we extract the constant parts from the form factors
for the fermion two-point function and the interaction terms to find the
flow equations for the couplings.  In this process, we extract the Z
factors out of the quantum corrections.

Inclusion of $\Gamma_{e^2}^{AA}$ in the Ansatz \labelcref{eq:1PI_for_flow} allows us
to find the moment dependent quantum correction that is consistent with
the flow equation and the QME in the lowest order in the coupling $e^2$.

\subsection{Quantum corrections to photon two-point function}
\label{q-corrections to photon 2 point}

We compute ${\cal A}_{\mu\nu}$ in \labelcref{eq:cal_A_again} by solving
the dimensionless flow equations \labelcref{eq:wetterich} with a
specific cutoff function $K(x) = \exp(-x)\,$ where $x= {p}^{2}$.  In
this subsection we describe only the results and leave further details
in the \Cref{app:PhotonProp}.

From the functional flow equation \labelcref{eq:wetterich}, we collect
terms with two photon fields $A_\mu A_\nu$ and obtain the flow equation
for ${\cal A}_{\mu\nu}$.  \Cref{fig:2pt photon} shows the one-loop
contribution on the r.h.s. of \labelcref{eq:wetterich}.  The $x$-linear
terms in the flow equation for ${\cal A}_{\mu\nu}$ have particular
roles: they give rise an algebraic relation \labelcref{eq:eta-A} for the
anomalous dimensions and the flow equation for the gauge parameter to be
presented shortly.  Similarly, from the flow equation for the fermion
two-point function, we find that $\Slash{p}$-linear terms give rise to
another relation \labelcref{eq:eta-psi} of anomalous dimensions.

The flow equation for ${\cal A}_{\mu\nu}$ leads to the differential
equations for ${\cal T}$ and ${\cal L}$ : at the lowest order in the
coupling $\alpha (t) = e^2(t)$, they are
\begin{align} \nonumber 
 \Bigl(x \frac{\pa }{\pa x}- 1\Bigr)
{\cal T}(t, x)  =&\, -\frac{\al}{8\pi^{2}x^{2}}\Bigl\{4 + 
\frac{2x^{3}}{3} - 
\bigl(4 + 2x - x^{2}\bigr)\exp(-x/2) 
 \Bigr\} \,,
 \\[2ex]
 \Bigl(x \frac{\pa }{\pa x}- 1\Bigr)
{\cal L}(t,x) =&\, \frac{\al}{8\pi^{2}x^{2}}
\Bigl\{12 -8 x - \bigl(12 - 2x - x^{2}
\bigr)\exp(-x/2) 
\Bigr\}~.
\label{eq:DiffT+L}
\end{align}
Solving \labelcref{eq:DiffT+L}, we obtain  
\begin{align}\nonumber 
 {\cal T}(t, x) =&\, \frac{\al(t)}{6\pi^{2}x^{2}}
\biggl[
1- \Bigl(1 + \frac{x}{2} -x^{2}\Bigr)\exp(-x/2)
+ \frac{x^{3}}{2} \Bigl({\rm Ei}(-x/2) - \log x\Bigr) 
\biggr] + c_T x\,, \\[1ex]
 {\cal L}(t, x)=&\, -
 \frac{\al(t)}{2\pi^{2} x^{2}} \left[1 - x -
 \left(1 - \frac{x}{2}\right)\exp(- x/2) 
\right]+c_L x\,,
\label{eq:calT+L}
\end{align}
where ${\rm Ei}$ is the exponential integral defined as,
\begin{align}
{\rm Ei}(-x/2) = \int_\infty^{x/2}  \frac{e^{-u}}{u} du \,.
\label{eq:Ei}
\end{align}
Note that ${\cal T}$ and ${\cal L}$ are $t$-dependent quantities through
their dependence on $\al(t)$.  Here $c_T$ and $c_L$ are integration
constants.  Here it is worth pointing out that the
potentially divergent contributions in the diagram to define ${\cal
A}_{\mu\nu}$ are contained in the integration constants.  Presently we
find that those constants should take finite values and, therefore,
both $\cal T$ and $\cal L$ are finite functions.

The condition $\Sigma^{AC} = 0$ determines the longitudinal part
${\cal L}$ in the photon two-point functions ${\cal A}_{\mu\nu}$. In
momentum space, it is given as
\begin{align}
{\cal L}(t, x={p^2}) = -8 \al(t) \int_{{q}} K({p}+{q})
{\bar K}({q})\frac{({p}\cdot {q})}{{{p}}^{2} {{q}}^{2}}\,,
\label{eq:longitudinal}
\end{align}
that is \labelcref{eq:calT+L_QME} in its dimensionless form.  For our
choice $K(x) = \exp (-x)$, \labelcref{eq:longitudinal} gives ${\cal
L}(t, x)$ in \labelcref{eq:calT+L} with $c_L =0$
\cite{Igarashi:2016gcf}: the consistency with QME requires $c_{L}=0$.
Since we have already renormalized fields, $x$-linear term should be
absent in $\cal T$.  This condition gives us
\begin{align}
c_T = \frac{1}{12 \pi^2} \biggl( \frac{13}{12} - \gamma + \log 2 \biggr)\,
\label{CT}
\end{align}
where $\gamma$ is the Euler's constant.

As we mentioned earlier, the flow equation for the photon two-point
function, obtained from \labelcref{eq:wetterich}, contains $x$-linear
terms both in the transverse and longitudinal parts.  Those in the
longitudinal part give the flow equation for the gauge parameter
\begin{align}
 \partial_t \xi = - \xi \cdot \eta_A(\alpha,\xi) \,.
\label{gauge parameter flow}
\end{align}
The \Cref{app:PhotonProp} contains its derivation.  From
\labelcref{gauge parameter flow} it is obvious that the Landau gauge
stays in the gauge along a RG flow.

Two algebraic relations \labelcref{eq:eta-A} and \labelcref{eq:eta-psi}
are solved to give $\eta_{A}$ and $\eta_{\psi}$ as
\begin{align}
& \eta_{A}(\al, \xi) = 
\Bigl\{ \cfrac{\al}{6\pi^{2}} \bigl(1 -\al  I^{(2)} \bigr) + 
\al^2 I^{(3)}I^{(4)}
\Bigr\} 
\Bigl\{1- \al I^{(2)} - \al^2  I^{(1)}I^{(4)} \Bigr\}^{-1} \,,
\nn\\
 & \eta_{\psi}(\al, \xi) = \al \cdot \Bigl\{I^{(3)} + \cfrac{\al}{6\pi^{2}} I^{(1)}\Bigr\}
\Bigl\{1- \al I^{(2)} - \al^2  I^{(1)}I^{(4)} \Bigr\}^{-1} \,,
 \label{eq:anom}
\end{align}
where 
\begin{align}\nonumber 
 I^{(1)}(\al, \xi) = &\,\frac{1}{32\pi^{2}} \int_{0}^{\infty}dx xK{\bar K}\Bigl\{
3 x K' T^{2} - \xi 
\Bigl(xK' + 2{\bar K}\Bigr) 
L^{2}\Bigr\} \,,\\[1ex]\nonumber 
 I^{(2)}(\al, \xi) =&\, -\frac{1}{32\pi^{2}} \int_{0}^{\infty}dx {\bar K} \Bigl\{
3x K' ({\bar K}-K)T + \xi \Bigl[2K{\bar K} - x K' ({\bar K}-K)\Bigr]L \Bigr\} \,,
\\[1ex]\nonumber 
 I^{(3)}(\al, \xi) = &\,\frac{1}{16\pi^{2}}\int_{0}^{\infty}dx \Bigl\{
 x^{2} K'\Bigl[3xK' T^{2} - \xi \bigl(xK'+2{\bar K}\bigr)L^{2}\Bigr] \\[1ex]\nonumber 
 &  - x{\bar K}
\Bigl[3(K'+xK'') T + \xi (K'-x K'') L\Bigr] \Bigr\} \,,
\\[1ex]
I^{(4)} =&\, - \frac{1}{6 \pi^2} \int_0^\infty dx \frac{K{\bar K}}{x} 
\Bigl( {\bar K}+x K' + x^2 K'' \Bigr) \,.
\label{eq:etapsi3}
\end{align}
\Cref{eq:anom,eq:etapsi3} are written for a generic regularization
function $K(x)$.  Note that the integrands on r.h.s. of
\labelcref{eq:etapsi3}, except $I^{(4)}$, depend on $\alpha$ and $\xi$
through $T(\al, x)$ and $L(\al, x, \xi)$ which are defined as
\labelcref{eq:TL} with the quantum corrections \labelcref{eq:calT+L}.

As a result of the renormalization, we obtained non-perturbative
expressions of the anomalous dimensions \labelcref{eq:anom} to be used
for flow equations.  In the lowest order in $\al$, \labelcref{eq:anom}
become
\begin{align}
\eta_A = \frac{\al}{6 \pi^2}\,,\qquad \eta_\psi = \al I^{(3)}(0,\xi) = \frac{\al}{8 \pi^2} \xi \,, 
\label{perturbative eta}
\end{align}
which are consistent with the known perturbative calculations.

\subsection{Flows of couplings} 
\label{Flows of couplings} 

The RG-improved one-loop diagrams that contribute to the flow of the
gauge coupling may be obtained in reference to \Cref{fig:q-corrections
to 3pt}.  The regulator function $\otimes = \partial_t {\cal R}-\eta
{\cal R}$ in \labelcref{eq:wetterich} is to be inserted to each of the
internal lines in \Cref{fig:q-corrections to 3pt}.  For example,
$e^3$-contribution consists of three diagrams shown in \Cref{fig:flow of
alpha with regulator insertion}.
Similarly we take account of diagrams with the regulator function
inserted in the other $eG$-diagram in \Cref{eG diagram}.  The diagrams
that contribute to the flow of the four-Fermi couplings are similarly
constructed from \Cref{fig:Gamma-4fermi-q}.

\begin{figure}[t]
  \begin{minipage}[b]{0.31\linewidth}
    \centering
   \includegraphics[keepaspectratio, scale=0.33]{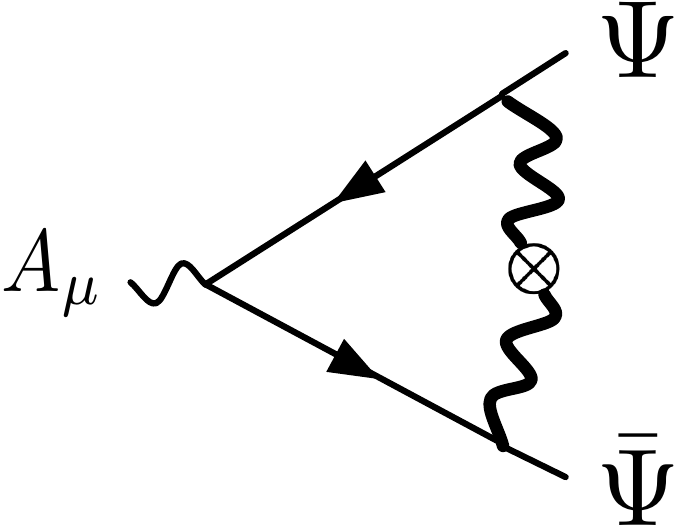}
  \end{minipage}
  \begin{minipage}[b]{0.33\linewidth}
    \centering
    \includegraphics[keepaspectratio, scale=0.33]{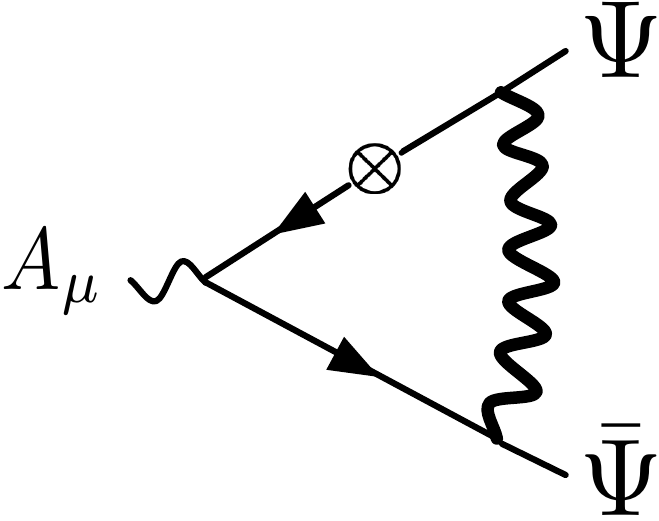}
  \end{minipage}
  \begin{minipage}[b]{0.32\linewidth}
    \centering
    \includegraphics[keepaspectratio, scale=0.33]{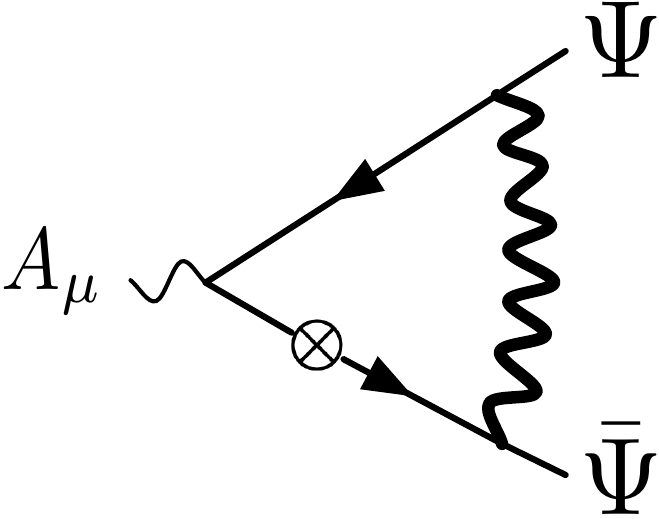}
  \end{minipage}
  \caption{The regulator denoted by $\otimes$ is inserted in each internal line.  All three diagrams contribute to the flow equation for $\al$.}
\label{fig:flow of alpha with regulator insertion}
\end{figure}

The quantum corrections to the photon propagators are taken up including
the momentum-dependent parts.  For the fermion, we only include quantum
corrections proportional to the free propagator that generate the wave
function renormalization $Z_2$.  In writing the dimensionless flow
equation given in \labelcref{eq:wetterich}, we extracted out $Z_2$ and
the internal fermion lines become the free propagators.
       
Finally we obtain the following flow equations for the couplings,
\begin{align}\nonumber 
 \pa_t \alpha = & \bigl(\eta_{A} + 2\eta_{\psi}\bigr)\al 
+ \frac{\al \bigl(G_{S} -4 G_{V}\bigr) }{4\pi^{2}} \int_0^\infty dx ~ K{\bar K} {\cal K}_\psi \\[1ex]\nonumber 
& \hspace{2cm} + \frac{\al^{2}}{8\pi^{2}} \xi \int_{0}^{\infty} dx ~
 K{\bar K}^{2} 
\bigl( x {\cal K}_{A} L^2 + 2 {\cal K}_{\psi} L \bigr) \,,
\\[2ex]\nonumber 
\partial_t G_{S}
 =& 2(1+ \eta_{\psi}) G_{S}
+ \frac{\bigl(3G_{S} -8 G_{V}\bigr) G_S }{8\pi^{2}} \int_0^\infty dx K{\bar K} {\cal K}_\psi \\[1ex]\nonumber 
& \hspace{2cm}
+ 
\frac{\al G_S}{8 \pi^2} \int_0^\infty dx K{\bar K}^2 \biggl(
x {\cal K}_A (3 T^2+\xi L^2) + 2 {\cal K}_\psi (3T+\xi L) \biggr) \\[1ex]\nonumber 
& \hspace{2cm}+ 
\frac{3 \al^2}{8 \pi^2} \int_0^\infty dx K{\bar K}^3 
\bigl( x {\cal K}_A T^3 + {\cal K}_\psi T^2 \bigr)\,,\\[2ex]\nonumber 
\partial_t G_{V}
 = & 2(1+ \eta_{\psi}) G_{V} 
- \frac{G_S^2}{16\pi^{2}} \int_0^\infty dx K{\bar K} {\cal K}_\psi \\[1ex]\nonumber 
& \hspace{2cm}
- 
\frac{\al G_V}{8 \pi^2} \int_0^\infty dx K{\bar K}^2 \biggl(
x {\cal K}_A (3 T^2-\xi L^2) + 2 {\cal K}_\psi (3T-\xi L) \biggr) \\[1ex]
& \hspace{2cm}
+ 
\frac{3 \al^2}{16 \pi^2} \int_0^\infty dx K{\bar K}^3 
\bigl( x {\cal K}_A T^3 + {\cal K}_\psi T^2 \bigr)\,,
\label{eq:flows}
\end{align}
where 
\begin{align}
{\cal K}_{A(\psi)}(x; \al, \xi) \equiv \eta_{A(\psi)}(\al, \xi) {\bar K}(x) + 2x K'(x)/K(x) \,.
\label{eq:calK}
\end{align} 
Again, the equations in \labelcref{eq:flows} are written for a generic
regularization function $K(x)$.  For our numerical calculations
to be explained shortly, we choose $K(x)=e^{-x}$.

In the last \Cref{q-corrections to photon 2 point}, we obtained the
photon two-point function as \labelcref{eq:calT+L} and
\labelcref{eq:longitudinal} from the flow equation
\labelcref{eq:wetterich} at the lowest order of the gauge coupling.  The
results are consistent with the requirement $\Sigma^{AC}=0$.

In \Cref{sec:barPsi_Psi_C}, we observed that the condition
$\Sigma^{\bar\Psi \Psi C}=0$ relates the quantum corrections
$\Gamma_{\rm q}^{\bar\Psi \Psi}$ and $\Gamma_{\rm q}^{\bar\Psi A \Psi}$
with their momentum dependence.  However, the Ansatz
\labelcref{eq:1PI_for_flow} does not allow the momentum-dependent
functions $\Gamma_{\rm q}^{\bar\Psi \Psi}$ and $\Gamma_{\rm q}^{\bar\Psi
A \Psi}$ and the flow equation with the Ansatz picks up only their
momentum independent contributions to give rise to the wave function
renormalization for the fermion and the flow equations for couplings
\labelcref{eq:flows}.  Therefore, with the present Ansatz, we do not
have the same condition $\Sigma^{\bar\Psi \Psi C}=0$ as in
\Cref{sec:barPsi_Psi_C}.

\subsection{The Hierarchical Phase Structure}

\begin{figure}[t]
  \begin{minipage}[b]{0.5\linewidth}
    \centering
    \includegraphics[keepaspectratio, scale=0.36]{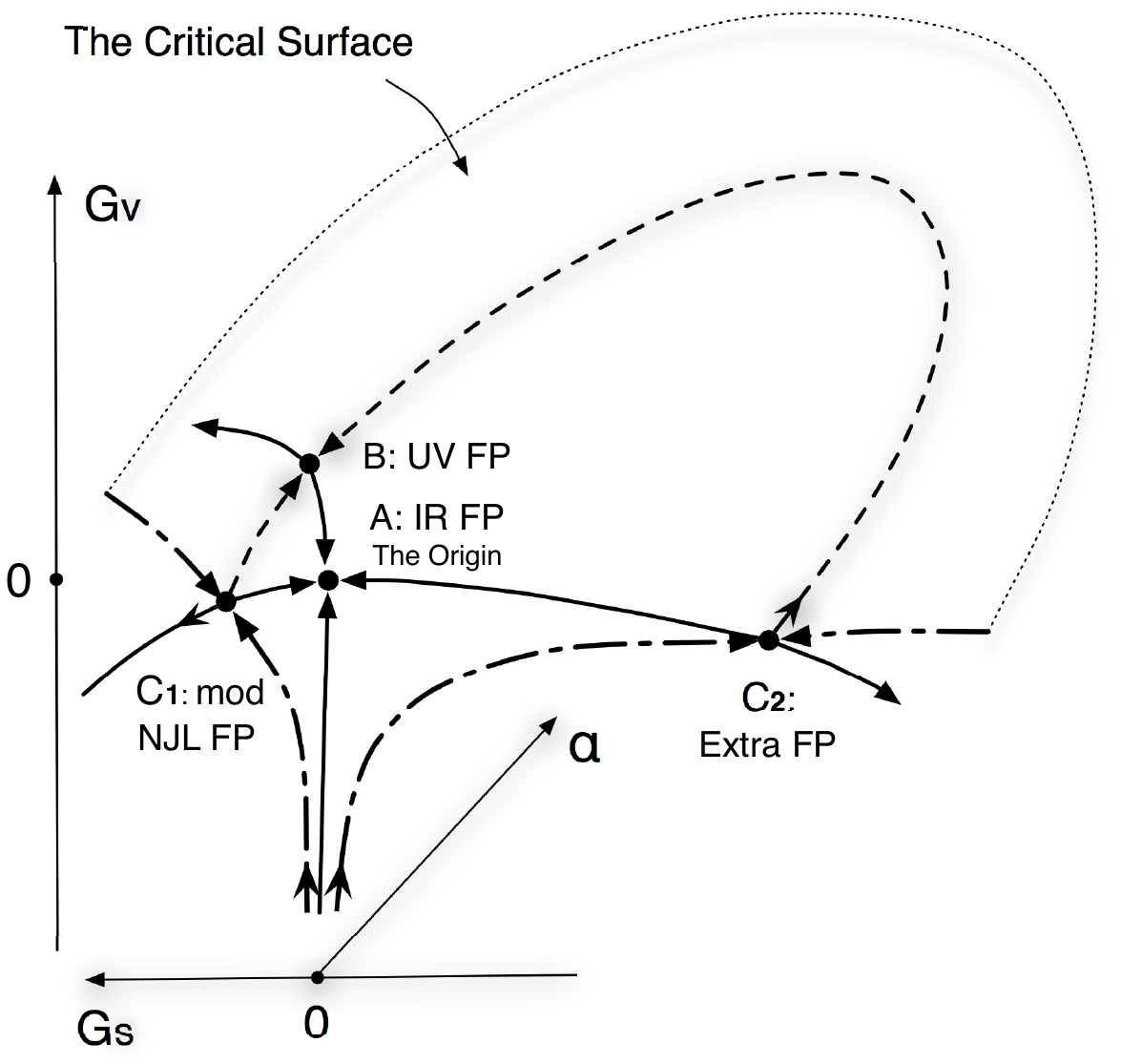}
     \caption{Fixed points and critical surface and critical lines.  Three
 fixed points, $A$, $C_1$ and $C_2$ are located on the $\alpha=0$
 plane.  The UV fixed point B has a non-zero value of $\alpha \sim 13.5$~.}  \label{phase structure}
  \end{minipage}
  \begin{minipage}[b]{0.4\linewidth}
    \centering
    \includegraphics[keepaspectratio, scale=0.45]{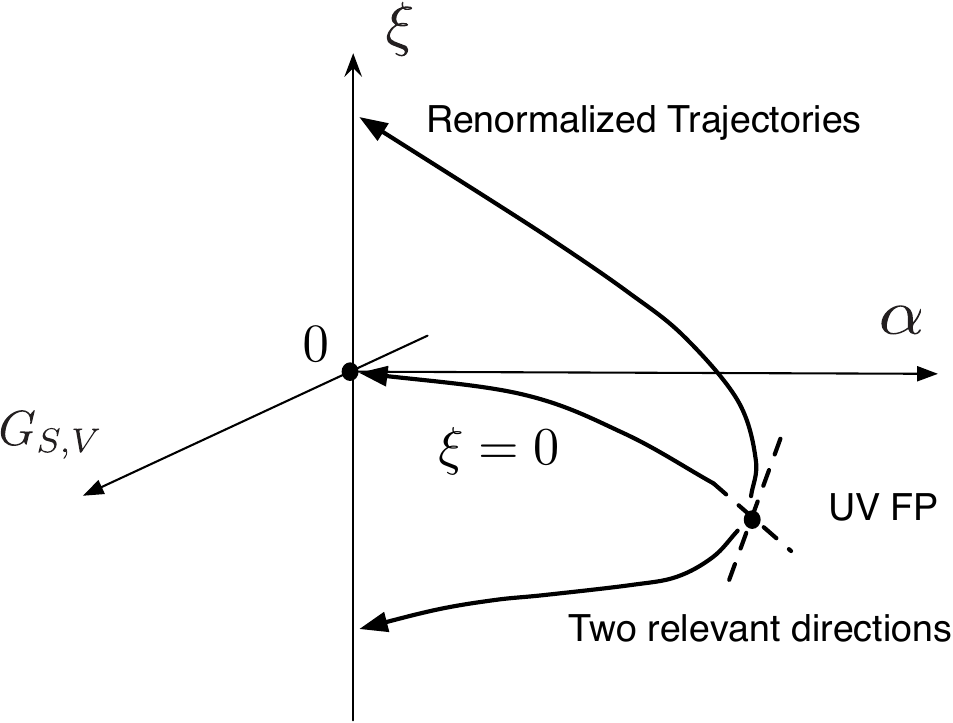}
    \caption{Flows including that of $\xi$. There appears an extra relevant
direction at the UV fixed point that generates flows in general
gauges.} \label{fig:Flowxi}
  \end{minipage}
\end{figure}

\begin{table}[b]
\center
  \begin{tabular}{lccc}
    Fixed Point&  ($\alpha, G_S, G_V$) & \# of Rel. Ops. \\ \hline
    A: IR & The origin & 0 \\
    B: UV & (13.5, 6.99, 0.573) & 1 \\
    $C_1$: mod. NJL & (0, 26.3, -3.29) & 2 \\
    $C_2$: extra & (0, -105, -52.6) & 2 \\ \hline
  \end{tabular}
\caption{Properties of four fixed points in the Landau gauge.}
\label{tab:FPs}
\end{table}
We first present our numerical results in the Landau gauge.
\Cref{phase structure} shows the phase structure obtained from the
flow equation \labelcref{eq:flows}.  Four fixed points are found: their
properties are listed on the table
\Cref{tab:FPs}.

Three fixed points, $A$, $C_1$ and $C_2$, are located on the $\alpha=0$
plane.  The point A is the origin, the IR fixed point.  $C_1$ is the
NJL fixed point modified due to the included vector four-Fermi coupling
$G_V$.  $C_2$ is usually regarded as an artifact owing to the truncation
of the action and ignored as an unphysical fixed point.  We view 
as the extra fixed point and included it since the hierarchical phase
structure is better viewed with its presence.  The fourth and most
important fixed point B is found with $\alpha \ne 0$.

The structure of flows around A and $C_1$ in the $\alpha = 0$ plane is well known,
and those around $C_2$ turned out to be copies of those around $C_1$.
 
We proceed with a discussion of the three-dimensional phase structure.
The point A is the IR fixed point.  In general, the boundary of its
realm forms a two-dimensional surface, the critical surface.  It was
found to be in the shape of a funnel halved by the $\alpha =0$ plane.
The densely dotted line in \Cref{phase structure} is drawn only for the
purpose of clearly showing the halved funnel. Flows restricted on the
critical surface run toward the fixed point B, which has one relevant
direction flowing out of the critical surface. The realm of the fixed
point B for flows on the critical surface has its one-dimensional
boundaries.  They are sections of the critical surface shaped in a
funnel by the $\alpha=0$ plane, i.e., two dash-dotted lines passing
through $C_1$ and $C_2$.

All the flows on these two lines go
into either $C_1$ or $C_2$. In three-dimensional space, $C_1$ and $C_2$
are fixed points with two relevant directions.

Here we clearly observe a hierarchical phase structure, and the flow
from B to A is the renormalized trajectory (RT).

In the last column of the Table \ref{tab:FPs}, we list the eigenvalues
for the relevant operators.  They take negative values due to our choice
of the parameter $t=\ln \Lambda / \mu$.

The order $O(\alpha)$ correction $\Gamma^{AA}_{e^2}$ contributes to the
photon propagator as \labelcref{eq:TL} that appears on r.h.s. of the flow
equations in \labelcref{eq:flows}.  Without the quantum correction
$\Gamma^{AA}_{e^2}$, the functions $T$ and $L$ are simply $1/x$ and
the UV fixed point is found at $(\alpha, G_S, G_V)=(13.8, ~8.17,~%
0.578)$: the values of $\alpha$ and $G_S$ differ approximately 2.5 $\%$
and 10 $\%$ from those on the Table \ref{tab:FPs}.

We also studied flows of couplings in \labelcref{eq:flows} together with
the flow equation for the gauge parameter $\xi$ given as
\labelcref{gauge parameter flow}.  By solving four flow equations, we
found that the point B on Table \ref{tab:FPs} is the unique UV fixed
point.  The linear analysis around B including \labelcref{gauge
parameter flow} gives us two relevant directions around the UV fixed
point.  Numerically solving the four flow equations, we understand that
the new relevant direction produces flows for gauges with $\xi \ne 0$ as
depicted in \Cref{fig:Flowxi}.

\subsection{Anomalous mass dimension at the UV fixed point}

\begin{figure}[t]
\center
 \includegraphics[scale=0.5]{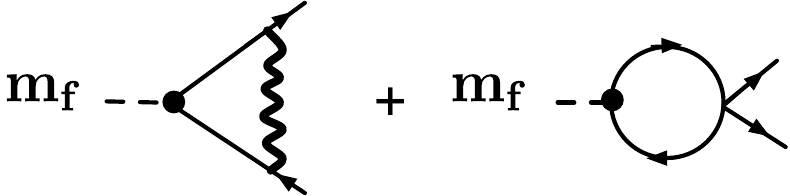}
\caption{Contributions to the mass anomalous dimension.  Blobs indicate
$m_f {\bar \psi}\psi$ operator.  To obtain $\gamma_m$, the regulator
$\partial {\cal R}-\eta {\cal R}$ in \labelcref{eq:wetterich} should be
inserted to internal propagators in the above diagrams.}
\label{AnomalousMassdim}
  \end{figure}
The renormalization flow equations for the present system was studied
earlier in \cite{Aoki:1996fh} with other regularization functions and
approximation.  The UV fixed point is identified in accordance with
works using Swinger-Dyson equation. Here we would like to calculate the
anomalous mass dimension and compare the result to that obtained in
\cite{Aoki:1996fh}.

Adding the mass operator $m_f {\bar \psi}\psi$ to $\Gamma_{I,\Lambda}$,
we define the anomalous mass dimension $\gamma_m$ by the following flow
equation,
\begin{align}
\partial_t m_f = - (1+\gamma_m) m_f + O(m_f^3)\,.
\label{eq:anomalous mass dimension}
\end{align}
In calculating $\gamma_m$ with the flow equation
\labelcref{eq:wetterich}, contributions are found from two diagrams in
\Cref{AnomalousMassdim}.  In evaluating the diagrams, we insert the
regulator $\partial {\cal R}-\eta {\cal R}$ in \labelcref{eq:wetterich}
to each one of internal propagators and sum up all possible
insertions. We obtain
\begin{align}\nonumber 
 \gamma_m (\alpha,G_S,\xi) =&\, - \eta_\psi + \frac{3 G_S}{4 \pi^2} \bigl(1-\frac{2}{9}\eta_\psi \bigr) \\[1ex]
 & - \frac{\alpha}{(4 \pi)^2} \int_0^\infty dx K{\bar K}^2 \Bigl[
  x {\cal K}_A (3 T^2 +\xi L^2) + 2 {\cal K}_\psi (3 T + \xi L)
  \Bigr] \,.
\label{eq:AnomMassDim}
\end{align}
Note that $\gamma_m$ in \labelcref{eq:AnomMassDim} does not depend on
$G_V$. Evaluating \labelcref{eq:AnomMassDim} at the UV fixed point, we
find $\gamma_m =$ 1.078, which is to be compared to 1.177 in
\cite{Aoki:1996fh}.  At the modified NJL fixed point, $G_S^*(\alpha=0) =
{8 \pi^2}/{3}$ and $\gamma_m(\alpha=0)=2$ in agreement with that of
\cite{Aoki:1996fh}.

\section{Gauge consistency}
\label{sec:Gauge_consistency}

In \Cref{sec:1PI+QME}, we have shown the QMEs 
$\Sigma^{AC} = \Sigma^{\bar\Psi\Psi C} =0$ for constant gauge and
four-Fermi couplings. These are results of the cancellation between the
antibracket terms and the measure terms in \labelcref{eq:Sigma_2} and
\labelcref{Sigma3e}.

\subsection{QME and renormalization}
\label{eq:QME+Ren}

In this subsection, we consider the QME in terms of the renormalized
quantities.  The key observation here is the fact the couplings of
$\Gamma_e$ in \Cref{eq:Gamma_e} behave differently under
renormalization.  Note that there are no genuine quantum corrections to
the second and third terms of $\Gamma_e$ and the renormalization of the
ghost is the only source of the change of those terms. In terms of
renormalized quantities $\Gamma_e$ now becomes
\begin{align} 
\Gamma_{e} = -e_{(r)}{\bar\Psi}_{(r)} \Slash{A}_{(r)}\Psi_{(r)} -i e'_{(r)}
\Psi_{(r)}^{*}\Psi_{(r)} C_{(r)} +i e'_{(r)}{\bar\Psi}_{(r)}^{*}\bar\Psi_{(r)} C_{(r)} \,,
\label{eq:Gamma10}
\end{align}
where
\begin{align} 
 e'_{(r)} = Z_{3}^{-1/2} e\,.
\label{eq:er-prime}
\end{align}
The difference of two couplings is given as
\begin{align} 
 e_{(r)} -e'_{(r)} = e_{(r)} [1- (Z_{e}Z_{3}^{1/2})^{-1}]\,.
\label{eq:Diff2Couplings}
\end{align}
It has been shown in  \cite{Igarashi:2021zml}, that  
$(Z_{e}Z_{3}^{1/2})-1 \propto G_{(r)S}-4G_{(r)V}$ in the one-loop
perturbative calculation.  Including contributions in higher order in
$e_{(r)}$, we are led to $(Z_{e}Z_{3}^{1/2})-1 \sim {\cal O}(e^2_{(r)},
G_{(r)S,V})$.

Let us find the implications of the above observation. We have shown in
\Cref{sec:1PI+QME}, that $\Sigma^{\bar\Psi\Psi C}_{e},
~\Sigma^{AC}_{e^{2}}, ~\Sigma^{\bar\Psi\Psi C}_{e^{3}}$ and
$\Sigma^{\bar\Psi\Psi C}_{e{\rm G}}$ vanish. Revisiting these
calculations, we notice that $e'_{(r)}$ could appear in measure terms as
well as the transformation of the fermion. Thus we conclude that all the
four $\Sigma$ listed above are proportional to $e_{(r)} -e'_{(r)}$ by
repeating the calculations in \Cref{sec:1PI+QME} using
\labelcref{eq:Gamma10}.  Here, as the simplest example, we take the BRST
transformation of the tree-level action $\bar\Psi_{(r)} (i
\Slash{\partial} -e_{(r)} \Slash{A}_{(r)}) \Psi_{(r)}$ and find
\begin{align}
\Sigma^{\bar\Psi\Psi C}_{e} = ie_{(r)} 
[1-( Z_{e}Z_{3}^{1/2})^{-1}]\int_{p,q}\bar\Psi_{(r)}(-q)
(\Slash{p}-\Slash{q})C_{(r)}(q-p)\Psi_{(r)}(p)\,.
\label{eq:e-e'-Simplest}
\end{align}
Similarly, we find
\begin{align}
\Sigma_{e^2}^{AC} \propto e_{(r)} ( e_{(r)} -e'_{(r)} )\,,
\qquad
\Sigma^{\bar\Psi\Psi C}_{e^3} \propto {e_{(r)}^{2}(e_{(r)} - e'_{(r)})}\,,
\qquad
\Sigma^{\bar\Psi\Psi C}_{eG} \propto {G_{(r)S,V}(e_{(r)} - e'_{(r)})}\,.
\label{eq:e-e'}
\end{align}

In the one-loop perturbative calculation, $e_{(r)} = e'_{(r)}$ in the
absence of the four-Fermi couplings and all the $\Sigma$ vanishes after
written with renormalized quantities.  Even in the presence of four-Fermi
couplings, each of $\Sigma$ listed above in eqs. \labelcref{eq:e-e'-Simplest}
and \labelcref{eq:e-e'} vanishes in the lowest order in couplings.

One may consider the possibility that the higher order terms due to the
difference of $e_{(r)}$ and $e'_{(r)}$ may cancel each other in the sum
$\Sigma^{\bar\Psi\Psi C}_{e} + \Sigma^{\bar\Psi\Psi C}_{e^{3}} +
~\Sigma^{\bar\Psi\Psi C}_{e {\rm G}}$.  However each term of the sum has
a distinctive integral form owing to its Feynman graphical origin and the
expected cancellation seems unlikely.  This also tells us that we cannot
extract any constraints among the couplings from the above sum of
$\Sigma$.

We close this discussion by summarizing our findings: we have shown in
\Cref{sec:1PI+QME} that $\Sigma^{AC}$ and $\Sigma^{\bar\Psi \Psi C}$
vanish with the RG-improved one-loop contributions to the 1PI effective
action. This does not produce any constraints among the couplings. When
repeating the calculations using the one-loop renormalized quantities,
$\Sigma^{AC}$ and $\Sigma^{\bar\Psi \Psi C}$ vanish if we consider only
the gauge coupling. In the presence of the four-Fermi couplings, they
vanish only in the lowest power of the couplings and the higher order
corrections remain.  Though we naturally expect even these higher order
corrections to $\Sigma^{AC}$ and $\Sigma^{\bar\Psi \Psi C}$ to vanish,
further work is required for a resolution of this intricacy.

\section{Summary and Discussions}
\label{sec:Sum+Dis}

We chose our 1PI effective action \labelcref{eq:mod.strlog}, as a
non-perturbative extension of \cite{Igarashi:2021zml}, based on the loop
expansion in referring to the flow equation.  For its construction, we
made an observation that the measure contribution to QME only has
fermion loops owing to linear gauge symmetry in QED: the vertices
$\Gamma^{(2)}_{I*}$ in \labelcref{eq:Sigma_K} have spinor indices only
as seen in the \Cref{app:vertices}.  This structure allows us to replace
the IR-regulated free photon propagator by a generic one in the QME.
For the consistency with the flow equation, we took the photon
propagator $\bar\Delta_{\mu\nu}$ with the lowest order correction
$\Gamma_{e^2}^{AA}$ to the photon two-point function. The quantum part
of our 1PI effective action consists of an expansion of ${\rm Str} \log
[(\Gamma_{\rm cl} + \Gamma_{\rm q}^{AA})^{(2)} + {\cal R}]$, the
RG-improved one-loop formula.  That is an extension of the one-loop
formula ${\rm Str} \log [\Gamma_{\rm cl}^{(2)} + {\cal R}]$ which
generates $\Gamma_{e^2}^{AA}$.  The 1PI effective action constructed in
this way simultaneously solve the flow equation and QME. Note that the
exact cancellation between the antibracket term and measure term in the
QME occur for the constant couplings.

Secondly we introduced wave function renormalization factors $Z$ and
running couplings to obtain the flow equations with anomalous
dimensions.  We noticed that special attention was necessary in writing
the QME in terms of the renormalized quantities.

The same constant coupling $e$ appears both in the gauge interaction and
the BRST transformation of the fermion fields.  If we could replace $e$
by the renormalized one $e_{r}=Z_{e} e$ in both terms, the QME holds
trivially.  However, with the one-loop perturbative calculation, we
found that the two terms are renormalized differently: the BRST
transformation of the fermion field acquires the coefficient
$e'_{r}=e_r/(Z_e Z_3^{1/2})$.  As discussed in
\Cref{sec:Gauge_consistency}, this makes the quantum master functional
proportional to $e_r- e'_r$.  After the one-loop perturbative
renormalization, the QME holds only approximately.  We leave this
question for a future study.

Thirdly, we reported our numerical study.  Our Ansatz contains the
quantum correction to the photon two-point function that is constructed
consistently with both the flow equation and QME.  Based on our
observation of QME, we used the dressed photon propagator
\labelcref{eq:phton_prop2} which is non-perturbative by its
construction.  The flow equation gave us the anomalous dimensions as
\labelcref{eq:anom} that is also non-perturbative and fully utilized in
our numerical study.  With our Ansatz and approximations, we identified
the UV fixed point whose properties are consistent with earlier works.

As for the quantum corrections to the fermion two-point function and the
three-point interaction vertex, only those proportional to their tree
forms are included in our numerical study: those quantum corrections
give rise to the renormalization of the gauge coupling and the anomalous
dimension for the fermion.  Incorporating form factors similarly to the
photon two-point function, we may be able to obtain the fermion
two-point function and the three-point interaction vertex discussed in
\Cref{sec:1PI+QME}.  Inclusion of those momentum dependent quantum
corrections is left for future research.

\section*{Acknowledgment}

Y. Igarashi and K. Itoh are grateful to J.-I. Sumi for stimulating
discussions. Y. Takahashi is grateful to M. Tanimoto and H. Nakano for
their encouragements.  The work by Y. Igarashi and K. Itoh has been
supported in part by the Grants-in-Aid for Scientific Research
nos. 19K03822 and 23K03412 from Japan Society for the Promotion of
Science. This work is supported by EMMI and by ERC-AdG-290623. This work
is also funded by the Deutsche Forschungsgemeinschaft (DFG, German
Research Foundation) under Germany’s Excellence Strategy EXC 2181/1 -
390900948 (the Heidelberg STRUCTURES Excellence Cluster) and the
Collaborative Research Centre SFB 1225 (ISOQUANT). It is also supported
by EMMI.

\appendix
\section{Vertices}
\label{app:vertices}

To calculate the quantum corrections $\Gamma_{q}$ and the measure term
$\Sigma_{K}$, we need field dependent parts of $\Gamma_{I}^{(2)}$,
namely the vertices $\tau[\Phi]$ defined in
\labelcref{eq:decomposeGamma2}, and $\Gamma_{I*}^{(2)}$ in
\labelcref{eq:Gamma2_star}. From $\Gamma_{\rm e}$ in
\labelcref{eq:Gamma_e} we find the following vertices,
\begin{align}\nonumber 
\tau_{\hat\al\beta}^{(-\Slash{A})}(x,y) = &\,
\frac{\pa^{l} \pa^{r} \Gamma_{\rm e}}{\pa \bar\Psi_{\hat\al}
(x) \pa \Psi_{\beta}(y)}
= -e  (\Slash{A})_{\hat\al\beta}(x) \delta(x-y) \,,
\\[1ex]\nonumber 
\tau_{\al\hat\beta}^{(\Slash{A}^{T})}(x,y) = &\,
\frac{\pa^{l} \pa^{r} \Gamma_{e}}{\pa \Psi_{\al}(x) 
\pa \bar\Psi_{\hat\beta}(y)}
= + e  (\Slash{A}^{T})_{\al\hat\beta}(x) \delta(x-y) \,,
\\[1ex]\nonumber 
\tau_{\hat\al\mu}^{(-\gamma\Psi)}(x,y) = &\,
\frac{\pa^{l} \pa^{r} \Gamma_{\rm e}}{\pa \bar\Psi_{\hat\al}(x) \pa A_{\mu}(y)} 
= - e (\gamma_{\mu}\Psi)_{\hat\al}(x) \delta(x-y)
=  -\tau_{\mu\hat\al}^{(\gamma\Psi)}(x,y) \,,
\\[1ex]
\tau_{\mu\beta}^{(-\bar\Psi\gamma)}(x,y) = &\,
  \frac{\pa^{l} \pa^{r} \Gamma_{e}}{\pa 
A_{\mu}(x) \pa\Psi_{\beta}(y)} = - e (\bar\Psi\gamma_{\mu})_{\beta}(x) 
\delta(x-y) = -\tau_{\beta\mu}^{(\bar\Psi\gamma)}(x,y)  \,.
\label{eq:vertices}
\end{align}
Here the superscripts of $\tau$ indicate structures of vertices. 
Similarly from $\Gamma_{\rm G}^{(2)}$, we have
\begin{align}\nonumber 
\tau_{\hat\al\beta}^{(\bar\Psi\Psi)}(x,y) 
=&\, \left[
\frac{\pa^{l} \pa^{r} \Gamma_{\rm G}[\Phi]}{\pa \bar\Psi_{\hat\al}
(x) \pa \Psi_{\beta}(y)}\right]_{A=0} 
\\[1ex]\nonumber 
=& \frac{G_{S}}{\Lambda^{2}}\delta(x-y)\biggl\{\Bigl[
 \delta_{{\hat\al}\beta}\bigl({\bar\Psi}(x)\Psi(x) \bigr)
- (\gamma_{5})_{{\hat\al}\beta} \bigl({\bar\Psi}(x)\gamma_{5}
\Psi(x) \bigr)\Bigr] 
\\[1ex]\nonumber 
&
- \Bigl[{\bar\Psi}_{\beta}(x)\Psi_{\hat\al}(x) 
- ({\bar\Psi}\gamma_{5})_{\beta}(x)(\gamma_{5}\Psi)_{\hat\al}(x)
\Bigr]
\biggr\}
\\[1ex]\nonumber 
& + \frac{G_{V}}{\Lambda^{2}}\delta(x-y) \biggl\{
\Bigl[
(\gamma_{\mu})_{{\hat\al}\beta}\bigl({\bar\Psi}(x)
\gamma_{\mu}\Psi(x)\bigr) 
+(\gamma_{5}\gamma_{\mu})_{{\hat\al}\beta}
\bigl({\bar\Psi}(x)\gamma_{5}\gamma_{\mu}\Psi(x)\bigr)\Bigr]
\\[1ex]\nonumber 
&
- \Bigl[\bigl({\bar\Psi}(x)\gamma_{\mu})_{\beta}
(\gamma_{\mu}\Psi(x)\bigr)_{\hat\al} 
+ \bigl({\bar\Psi}(x)\gamma_{5}\gamma_{\mu})_{\beta}
(\gamma_{5}\gamma_{\mu}\Psi(x)\bigr)_{\hat\al}\Bigr]\biggr\}
\\[1ex]
=&\, - \bigl(\tau^{(\bar\Psi\Psi)}\bigr)^{T}_{\beta\hat\al}(y,x)\,,
\label{eq:4f_2}
\end{align}
\begin{align}\nonumber 
\tau_{\al\beta}^{(\bar\Psi\bar\Psi)}(x,y) 
=&\, 
\frac{\pa^{l} \pa^{r} \Gamma_{\rm G}}{\pa \Psi_{\al}
(x) \pa \Psi_{\beta}(y)} 
\\[1ex]\nonumber 
=&\, -G_{S}\delta(x-y)\Bigl[\bar\Psi_{\al}(x)\bar\Psi_{\beta}(x)
- (\bar\Psi\gamma_{5})_{\al}(x)(\bar\Psi\gamma_{5})_{\beta}(x)\Bigr]
\\[1ex]
&\,-G_{V}\delta(x-y)\Bigl[(\bar\Psi\gamma_{\mu})_{\al}(x)
(\bar\Psi\gamma_{\mu})_{\beta}(x)
+ (\bar\Psi\gamma_{5}\gamma_{\mu})_{\al}(x)
(\bar\Psi\gamma_{5}\gamma_{\mu})_{\beta}(x)\Bigr]\,,
\end{align}
and 
\begin{align}\nonumber 
\tau_{\hat\al\hat\beta}^{(\Psi\Psi)}(x,y) 
=& \,
\frac{\pa^{l} \pa^{r} \Gamma_{\rm G}}{\pa \bar\Psi_{\hat\al}
(x) \pa \bar\Psi_{\hat\beta}(y)} 
\\[1ex]\nonumber 
=&\, -G_{S}\delta(x-y)\Bigl[\Psi_{\hat\al}(x)\Psi_{\hat\beta}(x)
- (\gamma_{5}\Psi)_{\hat\al}(x)(\gamma_{5}\Psi)_{\hat\beta}(x)\Bigr]
\\[1ex]
&-G_{V}\delta(x-y)\Bigl[(\gamma_{\mu}\Psi)_{\hat\al}(x)
(\gamma_{\mu}\Psi)_{\hat\beta}(x)
+ (\gamma_{5}\gamma_{\mu}\Psi)_{\hat\al}(x)
(\gamma_{5}\gamma_{\mu}\Psi)_{\hat\beta}(x)\Bigr]\,.
\end{align}
We find $\Gamma_{I*}^{(2)}$ in \labelcref{eq:Gamma2_star} as
\begin{align}\nonumber  
\tau^{C}_{*\al\beta} = &\,\frac{\pa}{\pa \Psi_{\al}^{*}(x)}
\frac{\pa^{r}}{\pa \Psi_{\beta}(y)} \Gamma_{e} = + ie \delta_{\al\beta} 
C(x) \delta(x-y)  \,,
\\[2ex]
\tau^{C}_{*\hat\al\hat\beta} = &\,\frac{\pa}{\pa \bar\Psi_{\hat\al}^{*}(x)}
\frac{\pa^{r}}{\pa \bar\Psi_{\hat\beta}(y)} \Gamma_{e} = - ie 
\delta_{\hat\al\hat\beta} 
C(x) \delta(x-y)  \,.
\label{Gamma*}
\end{align}

\section{Photon two-point function and anomalous dimensions}
\label{app:PhotonProp}

We derive the flow equation for the photon two-point function in
\labelcref{eq:DiffT+L} and two relations of the anomalous dimensions to
be solved as \labelcref{eq:anom}.

\subsection{Flow equation for photon two-point function}
\label{app:FlowPhoton2}

The flow equations for 
${\cal A}_{\mu\nu} = P_{\mu\nu}^T {\cal T}(t,x) + P_{\mu\nu}^L {\cal L}(t,x)$ ($x=p^2$) are
\begin{align}\nonumber 
& \Bigl( \pa_t - \eta_A(t) \Bigr) {\cal T} -  \eta_A(t) x  - 2 \Bigl(x \partial_x-1\Bigr){\cal T} 
= - 2 \alpha(t) C_T \bigl(\eta_\psi(t),x\bigr) \,, 
\\[1ex]
& \Bigl( \pa_t - \eta_A(t) \Bigr) {\cal L} 
+ x \Bigl( {\dot \xi (t)^{-1}} - \eta_A(t) \xi (t)^{-1} \Bigr) 
- 2 \Bigl(x {\partial_x}-1\Bigr){\cal L}
= - 2 \alpha(t) C_L \bigl(\eta_\psi(t),x \bigr) 
\,,
\label{eq:FlowCalL+T}
\end{align}
where $x=p^2$.  The two coefficient functions on the r.h.s. of \labelcref{eq:FlowCalL+T} are given as
\begin{align}\nonumber 
C_T(\eta_\psi(t), x) = &\, \frac{4}{3} \int_k
\frac{\bigl(K {\cal K}_\psi\bigr)(k)}{k^2}
\cdot
\frac{{\bar K}(p+k)}{(p+k)^2} 
\Bigl(3(p \cdot k) + k^2 + \frac{2(p \cdot k)^2}{p^2} \Bigr) \,,   \\[1ex]
C_L(\eta_\psi(t), x) =&\,  -4 \int_k 
\frac{\bigl(K {\cal K}_\psi\bigr)(k)}{k^2}
\cdot
\frac{{\bar K}(p+k)}{(p+k)^2} 
\Bigl((p \cdot k) - k^2 + \frac{2(p \cdot k)^2}{p^2}
\Bigr) \,, \label{eq:CoeffFuncCs} 
\end{align}
where ${\cal K}_\psi$ is defined in \labelcref{eq:calK} as
\begin{align}
{\cal K}_{\psi}(\eta_{\psi}, ~x) \equiv \eta_{\psi} {\bar K}(x) + 2x K'(x)/K(x) \,.
\label{eq:calK2}
\end{align} 
$C_T$ and $C_L$ in \labelcref{eq:CoeffFuncCs} are functionals
of the regulator function $K$.  They depend on the scale $t$ through $\eta_\psi(t)$.

In \labelcref{eq:FlowCalL+T}, $x$-linear terms play special roles.  Let
us first consider \labelcref{eq:FlowCalL+T} for $\cal T$.  Due to
renormalization, $\cal T$ does not have any $x$-linear terms.  In
\labelcref{eq:FlowCalL+T}, the anomalous dimension on the l.h.s. must
balance to the coefficient of the $x$-linear term on the r.h.s.: we find
a relation of the anomalous dimensions
\begin{align}
\eta_A = \frac{\al}{6 \pi^2} + \al \eta_\psi I^{(4)}
\label{eq:eta-A}
\end{align}
where
\begin{align}
I^{(4)} = - \frac{1}{6 \pi^2} \int_0^\infty dx \frac{K{\bar K}}{x} 
\Bigl( {\bar K}+x K' + x^2 K'' \Bigr) \,.
\label{eq:I4}
\end{align}
In \labelcref{eq:FlowCalL+T} for $\cal L$, we note terms linear in $x$ and
the vanishing of its coefficient gives rise to the flow equation for the
gauge parameter
\begin{align}
{\dot \xi (t)^{-1}} = \eta_A(t) \xi (t)^{-1}\,.
\label{eq:Flowxi}
\end{align}
This is consistent with \labelcref{eq:couplings}.

It would be appropriate to explain why other terms of the second
equation of \labelcref{eq:FlowCalL+T} do not produce any $x$-linear
terms: (i) as we will find in \Cref{eq:calT+L-Behaviour}, $\cal L$
vanishes for a large $x$ and the first term on the l.h.s. does not have any
$x$-linear terms, at least in the lowest order in $\alpha$; (ii) in the
third term on the l.h.s., the operator $x \pa_x -1$ removes $x$-linear
terms in $\cal L$; (iii) we find that the coefficient function
$C_L(\eta_\psi(t), x)$ in \labelcref{eq:FlowCalL+T} does not have any
$x$-linear terms after some calculations.  The observation (ii) is
trivial and let us explain two other properties.

We start from the property (iii).  With the regulator function
$K(x)=e^{-x}$, we may explicitly calculate the coefficient function
$C_L$ in \labelcref{eq:CoeffFuncCs} as
\begin{align}\nonumber
C_L (\eta_\psi, x) =& \,\frac{3}{2 \pi^2 x^2} 
\Bigl\{1-\frac{2x}{3}-\Bigl(1-\frac{x}{6}-\frac{x^2}{12}\Bigr) e^{-x/2} \Bigr\}
\\[1ex]
& +
\frac{\eta_\psi}{4 \pi^2 x^2} 
\Bigl\{ - \frac{3}{4}\Bigl(x-\frac{7}{6}\Bigr) + (x-2) e^{-x/2} 
- \frac{1}{2}\Bigl(x-\frac{9}{4}\Bigr) e^{-2x/3}
\Bigr\} \,,
\label{eq:CLExp}
\end{align}
by using the integral representation of the modified Bessel function
\begin{align}
\int_0^{\pi} {\rm d} \theta \exp (2 p q \cos \theta) \sin^2 \theta 
= \frac{\pi}{2 p q} I_1 (2 p q) \,.
\label{modified Bessel}
\end{align}
One can confirm that there is no $x$-linear term in \labelcref{eq:CLExp}
and it is easy to see that $C_L (\eta_\psi, x)$ vanishes for a large
$x$.  We have not proved this property for a generic regulator function
$K(x)$.

Shortly we find the solution to \labelcref{eq:FlowCalL+T} in the lowest
order in $\alpha$ as given in \labelcref{eq:calT+L}.  The solution
vanishes as $1/x$ for a large $x$ (cf. \labelcref{eq:calT+L-Behaviour}).
That implies \labelcref{eq:Flowxi} since \labelcref{eq:FlowCalL+T} is
understood to hold for a generic $x$.

In the lowest order in $\al$, \labelcref{eq:FlowCalL+T} and
\labelcref{eq:FlowCalL+T} become
\begin{align}
 \Bigl(x \partial_x-1\Bigr){\cal T}(t ,x) 
=  \alpha C_T \bigl(0, x\bigr) 
-\frac{\al}{12 \pi^2}
\,, 
\qquad \Bigl(x {\partial_x}-1\Bigr){\cal L}(t ,x) 
= \alpha C_L \bigl(0, x \bigr) 
\,.
\label{FlowCalT+LLowest}
\end{align}
The coefficient functions of r.h.s. of \labelcref{FlowCalT+LLowest} are
\begin{align}\nonumber 
&
C_T(0, x) =  \frac{8}{3} \int_k
\frac{K'(k){\bar K}(p+k)}{(p+k)^2} 
\Bigl(3(p \cdot k) + k^2 + \frac{2(p \cdot k)^2}{p^2} \Bigr) \,,  \\[1ex]
&
C_L(0, x) =  -8 \int_k 
\frac{K'(k){\bar K}(p+k)}{(p+k)^2} 
\Bigl((p \cdot k) - k^2 + \frac{2(p \cdot k)^2}{p^2}
\Bigr) \,. 
\label{eq:CoefFuncCs-0102} 
\end{align}
For $K(x)= e^{-x}$, \labelcref{FlowCalT+LLowest} becomes  
\begin{align} 
& \Bigl(x \frac{\pa }{\pa x}- 1\Bigr)
{\cal T}(t, x)  = -\frac{\al}{8\pi^{2}x^{2}}\Bigl\{4 + 
\frac{2x^{3}}{3} - 
\bigl(4 + 2x - x^{2}\bigr)\exp(-x/2) 
 \Bigr\} \,,
 \nn\\
& \Bigl(x \frac{\pa }{\pa x}- 1\Bigr)
{\cal L}(t,x) = \frac{\al}{8\pi^{2}x^{2}}
\Bigl\{12 -8 x - \bigl(12 - 2x - x^{2}
\bigr)\exp(-x/2) 
\Bigr\}~.
\label{eq:DiffT+L2}
\end{align}
Solving \labelcref{eq:DiffT+L2}, we obtain  
\begin{align}
\nonumber 
 {\cal T}(t, x) =&\, \frac{\al(t)}{6\pi^{2}x^{2}}
\biggl[
1- \Bigl(1 + \frac{x}{2} -x^{2}\Bigr)\exp(-x/2)
+ \frac{x^{3}}{2} \Bigl( {\rm Ei}(-x/2) - \log x \Bigr)
\biggr] + c_T x~, \\[1ex]
& {\cal L}(t, x)= -
 \frac{\al(t)}{2\pi^{2} x^{2}} 
\biggl[ 1 - x -
 \left(1 - \frac{x}{2}\right)\exp(- x/2) 
\biggr] +c_L x\,,
\label{eq:calT+Lapp}
\end{align}
where ${\rm Ei}(-x/2)$ denotes the exponential integral defined as 
\begin{align}
{\rm Ei}(-y) = \int_\infty^{y}  \frac{e^{-u}}{u} du \,.
\label{eq:Ei2}
\end{align}
Note that ${\cal T}$ and ${\cal L}$ are $t$-dependent quantities through
their dependence on $\al(t)$.  Here $c_T$ and $c_L$ are integration
constants.  We will show presently that those constants are
\begin{align}
c_T = \frac{1}{12 \pi^2} \biggl( \frac{13}{12} - \gamma + \log 2 \biggr)\,,
\qquad 
c_L = 0 \,
\label{eq:IntConsts}
\end{align}
where $\gamma$ is the Euler's constant that appears in the power series
expansion of ${\rm Ei}(-x/2)$.

The transverse part $\cal T$ does not have any $x$-linear terms due to
the renormalization: the condition determines the above value of $c_T$.
Here we used the expansion
\begin{align}
{\rm Ei}(-y) = \gamma + \log y + \sum_{n=1}^{\infty} \frac{(-1)^n y^n}{n!~ n} \,.
\label{eq:ExpEi}
\end{align}
The QME $\Sigma^{AC} = 0$ can be used to determine the longitudinal part
${\cal L}$.  In momentum space, it is given as
\begin{align}
{\cal L}(t, x) = -8 \al(t) \int_{{q}} K({p}+{q})
{\bar K}({q})\frac{({p}\cdot {q})}{{{p}}^{2} {{q}}^{2}}\,.
\label{eq:longitudinal}
\end{align}
For our choice $K(x) = e^{-x}$, \labelcref{eq:longitudinal} gives ${\cal
L}(t, x)$ in \labelcref{eq:calT+L} with $c_L =0$
\cite{Igarashi:2016gcf}.  The consistency with QME requires $c_{L}=0$.

Two comments are in order. For large value of $x=p^2$, the exponential integral
${\rm Ei}(-x/2)$ tends to zero and we find
\begin{align}
{\cal T}(t,x) \sim -\frac{\alpha(t)}{6 \pi^2}\cdot \frac{1}{2} \cdot x \log x \,.
\label{eq:calT+L-Largex}
\end{align}
$\alpha(t)/(6 \pi^2)$ is the anomalous dimension of the photon in the lowest order.  
\labelcref{eq:calT+L-Largex} may be regarded as the lowest order expansion of
the following behaviour
\begin{align}
p^2 + {\cal T}(t,p^2) \sim p^{2-\eta_A}\,.
\end{align}
$\cal L$ in \labelcref{eq:calT+L} with $c_L=0$ behaves for small and large $x$ as 
\begin{align}
{\cal L}(t,x) \sim \frac{\alpha(t)}{2 \pi^2} \times 
\left\{
\begin{array}{ll}
3/8 - x/12 & x \sim 0 \\
1/x        & x \sim \infty \,.
\end{array}
\right.
\label{eq:calT+L-Behaviour}
\end{align}

\subsection{Relation between anomalous dimensions from fermion two-point function}
\label{app:RelAnomalous}

By repeating a similar calculation for the fermion two-point function,
we find the following relation among the anomalous dimensions
\begin{align}
\eta_\psi = \al \bigl(\eta_A I^{(1)} + \eta_\psi I^{(2)} + I^{(3)}\bigr)\,.
\label{eq:eta-psi}
\end{align}
Solving \labelcref{eq:eta-A} and \labelcref{eq:eta-psi} for $\eta_A$ and
$\eta_\psi$, we obtain \labelcref{eq:anom}.

\section{Gauge-consistent fRG expansion scheme}
\label{app:GenStructureFlow}

In this Appendix we augment the explicit derivations in this work with a
more structural point of view on our approach. This is specifically
important for potential systematic improvements of the approach. We may
split the workflow into two structural parts.
\begin{enumerate}
    \item Construct a gauge-consistent approximation
	  $\Gamma_\textrm{gc}$ to the effective action. This
	  approximations is given in terms of the effective action
	  itself and its derivatives, see
	  e.g.~\labelcref{eq:RG-improved} or \labelcref{eq:Skeleton2} in
	  \Cref{app:SystematicSkeleton} below.
    \item Derive flow equations for the gauge-consistent couplings of
	  $\Gamma_\textrm{gc}$ within a respective approximation of the
	  flow equation. For a structural point of view see
	  \Cref{app:RG-Improvement}.
\end{enumerate}
The second step will be discussed in \Cref{app:RG-Improvement} where we
show that the 1PI effective action with the the RG-improved formula
\labelcref{eq:RG-improved} can be understood as an approximate solution
to the flow equation with a particular set of interaction vertices. Put
differently, we show how the flow of the vertices of
\labelcref{eq:RG-improved} are computed from the flow equation.

\subsection{Systematic skeleton expansion scheme} 
\label{app:SystematicSkeleton}

We start our analysis with restructuring the Wetterich equation
\labelcref{eq:FlowGamma} as a total derivative and RG-improvement terms,
\begin{align} 
\partial_t \Gamma = \frac{1}{2}{\rm Str} \frac{1}{\Gamma^{(2)} + {\cal R} }\partial_t {\cal R} =\frac12 {\rm Str}\,\partial_t  \ln \left[\Gamma^{(2)} + {\cal R} \right] - \frac12 {\rm Str} \frac{1}{\Gamma^{(2)} + {\cal R} }\partial_t \Gamma^{(2)}\,.
\label{eq:Skeleton1}
\end{align} 
Upon $t$-integration from the initial cutoff scale $t_0$ to $t$, \labelcref{eq:Skeleton1} gives us, 
\begin{subequations}
\label{eq:Skeleton}
\begin{align} 
\Gamma (t) =\Gamma^{1}_{\textrm{sk}}(t)- \frac12 \int_{t_0}^t {d t^\prime}~{\rm Str} 
\frac{1}{(\Gamma^{(2)}+ {\cal R})(t^\prime) }\partial_{t^\prime} \Gamma^{(2)}(t^\prime)\,, 
\label{eq:Skeleton2}
\end{align} 
with the first order $\Gamma^{1}_{\textrm{sk}}(t)$ of the skeleton
expansion in the presence of the regulator,
\begin{align}  
\Gamma^{1}_{\textrm{sk}}(t) = \Gamma(t_0) + \frac12 {\rm Str}\left(\ln \left[\Gamma^{(2)} + {\cal R}\right](t)-\ln \left[\Gamma^{(2)} + {\cal R} \right](t_0)\right)\,, 
\label{eq:Skeleton3}
\end{align} 
\end{subequations} 
the effective action in \labelcref{eq:RG-improved}. The structure of
\labelcref{eq:Skeleton2} is that of a one-loop functional of the full
second derivative of the effective action, $\Gamma^{(2)}$, and an
RG-improvement term. This term can be rewritten as a two-loop functional
in terms of $\Gamma^{(2)}$ similarly to \labelcref{eq:Skeleton3}, and a
higher order RG-improvement term. For the sake of illustration we
indicate this procedure with the next order. In a first step we use the
skeleton expansion \labelcref{eq:Skeleton} for the flow of
$\Gamma^{(2)}$. This results in
\begin{align} 
\partial_t \Gamma^{(2)} =&\, \frac12 {\rm Str}\, \partial_t \left[\frac{1}{\Gamma^{(2)} + {\cal R}} \left(\Gamma^{(4)}-\Gamma^{(3)}\frac{1}{\Gamma^{(2)} + {\cal R}} \Gamma^{(3)} \right)\right]-\frac12 {\rm Str} \left[
\frac{1}{\Gamma^{(2)}+ {\cal R}}\partial_t \Gamma^{(2)}\right]^{(2)}\,. 
\label{eq:SkeletonGamma2}
\end{align} 

The last comprises second order terms in the skeleton expansion, which
only contribute to third order ones, if \labelcref{eq:SkeletonGamma2} is
inserted in the last term of \labelcref{eq:Skeleton2}: the give rise to
integrated two-loop terms similar to that in
\labelcref{eq:Skeleton2}. As $\partial_t \Gamma^{(2)}$ is given by a
skeleton loop itself, it is at least third order in the skeleton
expansion.

The first term on the right hand side of \labelcref{eq:SkeletonGamma2}
is a total derivative. If inserted into the integrand in the last term
in \labelcref{eq:Skeleton2}, the integrand can be written as a total
derivative and higher order terms. We find
\begin{align} \nonumber 
\frac12 {\rm Str} 
\frac{1}{\Gamma^{(2)}+ {\cal R} }\partial_t \Gamma^{(2)} = &\,\partial_{t}\left[\frac14   
\frac{1}{\Gamma^{(2)}+ {\cal R}  }\Gamma^{(4)}\frac{1}{\Gamma^{(2)}+ {\cal R} }- \frac16 \,    
\Gamma^{(3)}\left(\frac{1}{\Gamma^{(2)}+ {\cal R}  }\right)^3 \Gamma^{(3)}\right]\\ 
& + 3\textrm{rd-order terms}\,, 
\end{align}
where all propagators are fully contracted with the vertices. In the
second term all propagators are contracted with both vertices. Upon
integration the first term provides the second order skeleton terms and
higher order terms. For the second order terms we can choose a gauge
consistent approximation for $\Gamma^{(n)}$, thereby solving the QME. We
shall discuss this explicitly for the first order used in the present
work in \labelcref{app:RG-Improvement}.

Finally we remark that this procedure can be iterated systematically. In
summary, the general gauge-consistent procedure discussed in the present
work is now implemented as follows: first we observe that the QME
connects different orders in the skeleton expansion. Still we reduce the
\labelcref{eq:Skeleton2} to its first order contribution and implement a
gauge-consistent approximation to $\Gamma^{1}_{\textrm{sk}}$ in terms of
a respective approximation of $\Gamma^{(2)}$. Diagrams derived from
\labelcref{eq:Skeleton3} are the one-loop diagrams (lowest order) in
full propagators and vertices in a skeleton expansion. In
\Cref{app:RG-Improvement} we shall discuss the gauge consistency of
\labelcref{eq:Skeleton3} with an appropriate approximation for
$\Gamma^{(2)}$.

\subsection{Flow equation for the RG-improved action}
\label{app:RG-Improvement}

Before looking into the RG-improved formula in \labelcref{eq:RG-improved}, let
us first consider the standard one-loop approximation in
\labelcref{eq:Gammacl-1loop}.  This formula is obtained by putting only the
classical interaction action $\Gamma_{I,{\rm cl}}$ on the r.h.s. of the
flow equation at all the scale $t$: all the accumulating quantum
contributions are ignored at every step of solving the flow equation.
Similarly we will presently find that the RG-improved formula
\labelcref{eq:RG-improved} is obtained collecting particular set of quantum
corrections.

Suppose now we have the action with \labelcref{eq:RG-improved} added to
the classical action at a scale $t$:
\begin{align}
\Gamma_I(t) = \Gamma_{I, {\rm cl}} 
+ \frac{1}{2}{\rm Str} \log\,\Bigl[ (\Gamma_{\rm cl} + \Gamma_{\rm q}^{AA})^{(2)} + {\cal R}\Bigr](t)\,.
\label{Gamma I at t}
\end{align}
The second term on the r.h.s. of \labelcref{Gamma I at t} generate terms
in \labelcref{eq:mod.strlog2} with the classical vertices, the tree
fermion and the dressed photon propagators.

The condition that $\Gamma_I(t+\Delta t)$ generated as
\labelcref{eq:flow_Gamma_int2} is also given by the formula
\labelcref{Gamma I at t} is expressed as
\begin{align}
\frac{d}{dt} {\rm Str} \log\,\Bigl[ (\Gamma_{\rm cl} + \Gamma_{\rm q}^{AA})^{(2)} + {\cal R}\Bigr](t)\,
=\frac{1}{2}{\rm Str}\left[\otimes \frac{1}{\bar\Delta^{-1}  +  \tau[\Phi]}\right](t)\,.
\label{Condition on corrections}
\end{align}
${\bar \Delta}^{-1}$ and $\tau[\Phi]$ are defined at the sale $t$ as
eqs. \labelcref{eq:inversepropagator} and
\labelcref{eq:decomposeGamma2}.  \labelcref{Gamma I at t} satisfies the
flow equation with some approximations and we will explain how
\labelcref{Condition on corrections} holds by taking appropriate
vertices for $\tau[\Phi]$ on the r.h.s..

The expansion of the RG-improved formula is given in \labelcref{eq:mod.strlog2} as
\begin{align}
{\rm Str} \log\,\Bigl[ (\Gamma_{\rm cl} + \Gamma_{\rm q}^{AA})^{(2)} + {\cal R}\Bigr]
= \Gamma_{e^2}^{AA} + \Gamma_{e^2}^{{\bar \Psi}\Psi} 
+ \Gamma_{\rm q}^{{\bar \Psi} A \Psi} 
+ \Gamma_{\rm q}^{\bar\Psi \Psi\bar\Psi \Psi} + \cdots\,.
\label{mod.strlog2-2}
\end{align}
The first term $\Gamma^{AA}_{e^2}$ consists of a fermion one-loop.  For
this term \labelcref{Condition on corrections} is satisfied with the
classical vertices $\tau_{\rm cl}$, the 3-point vertices
$\tau^{(-\Slash{A})}$ and $\tau^{(\Slash{A}^T)}$ given in
\labelcref{eq:vertices}.  Similarly for terms in
\labelcref{mod.strlog2-2} only with fermion internal lines, we need the
classical vertices to satisfy \labelcref{Condition on corrections}.

The second term $\Gamma_{e^2}^{{\bar \Psi}\Psi}$ contains a fermion and
a dressed photon propagators.  Taking its derivative, we find
\begin{align}
{\dot\Gamma}_{\rm e^2}^{\bar\Psi\Psi}
= -e^{2} \bar\Psi \gamma_{\mu}
\Bigl(
\dot{\bar\Delta}^{(0)} \gamma_{\nu} {\bar\Delta}_{\mu\nu}
+{\bar\Delta}^{(0)} \gamma_{\nu}
\dot{\bar\Delta}_{\mu\nu}
\Bigr)
\Psi\,.
\label{pa_t fermi-2-point}
\end{align}
The derivative of the dressed photon propagator acts on $\cal A$ as well
as the regulator $\cal R$,

\begin{align}
\pa_{t} \bar\Delta_{\mu\nu} 
= - \left(\bar\Delta \pa_{t} \bigl[\bigl({\bar\Delta}^{(0)}\bigr)^{-1}+{\cal A}\bigr] \bar\Delta\right)_{\mu\nu}
= - \left( \bar\Delta  \bigl[ \dot{\cal R} + \dot{\cal A}    \bigr]  \bar\Delta  \right)_{\mu\nu}\,.
\label{pa_t phtonpro}
\end{align}
$\cal R$ is the regulator defined in \labelcref{eq:regulator} and its derivative is
\begin{align}
\dot {\cal R} = \otimes = -\bigl({\bar\Delta}^{(0)}\bigr)^{-1} {\dot{\bar\Delta}}^{(0)} \bigl({\bar\Delta}^{(0)}\bigr)^{-1}\,.
\label{dot cal R}
\end{align}
The flow of $\Gamma_{\rm e^2}^{\bar\Psi\Psi}$ is given by 
\begin{align}
{\dot\Gamma}_{\rm e^2}^{\bar\Psi\Psi}
=&\, e^{2} \bar\Psi \gamma_{\mu}
\Bigl[
\bigl( {\bar\Delta}^{(0)}  {\dot {\cal R}}   {\bar\Delta}^{(0)} \bigr)
\gamma_{\nu} {\bar\Delta}_{\mu\nu}
+ {\bar\Delta}^{(0)} \gamma_{\nu} 
\bigl( \bar\Delta  \dot{\cal R}  \bar\Delta  \bigr)_{\mu\nu}
\Bigr] \Psi\, 
\nn\\
&~ + ~e^{2} \Bigl(\bar\Psi \gamma_{\mu} 
{\bar\Delta}^{(0)} \gamma_{\nu} \Psi \Bigr)
\bigl( \bar\Delta   \dot{\cal A}  \bar\Delta \bigr)_{\mu\nu} \,.  
\label{pa_t fermi-2-point 2}
\end{align}
The terms on the first line of r.h.s. of \labelcref{pa_t fermi-2-point 2}
is in the expected RG-improved one-loop form with $\tau_{\rm cl}$, while
the last term is obviously in two-loop.  We will show that this term may
also be rewritten in the form of the l.h.s. of \labelcref{Condition on
corrections} with an appropriate vertex $\tau[\Phi]$.

\begin{figure}[H]
   \centering
   \includegraphics[width=0.45\columnwidth]{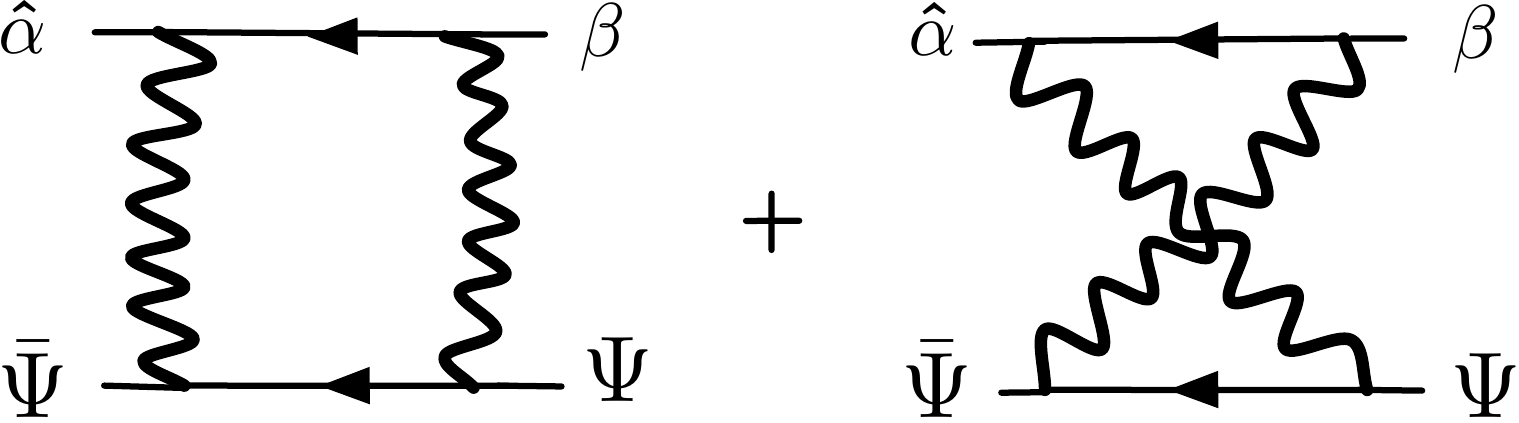}
\caption{$\bigl(\tau_{e^4}^{{\bar \Psi}\Psi}\bigr)_{{\hat \alpha} \beta}$ in \labelcref{2nd term b}.}
\label{fig:tau-e4}
\end{figure}

Using ${\cal A}_{\rho\sigma}$ given in \labelcref{eq:AA}
\begin{align}
{\cal A}_{\rho\sigma} = \frac{e^2}{2} {\rm tr} 
\Bigl( {\bar \Delta}^{(0)} \gamma_\rho {\bar \Delta}^{(0)} \gamma_\sigma   \Bigr)\,,
\label{cal A}
\end{align}
we find the second term in \labelcref{pa_t fermi-2-point 2} as
\begin{align}
&e^{2} \Bigl( \bar\Psi \gamma_{\mu} 
{\bar\Delta}^{(0)} \gamma_{\nu} \Psi \Bigr)
\bigl( \bar\Delta   \dot{\cal A}  \bar\Delta \bigr)_{\mu\nu}   
\nn\\
&~~=
e^2 \Bigl( {\bar \Psi} \gamma_\mu {\bar\Delta}^{(0)} \gamma_{\nu} \Psi \Bigr)
{\bar \Delta}_{\mu\rho} {\bar \Delta}_{\nu\sigma} \cdot 
\frac{e^2}{2} \Bigl[ 
{\rm tr} \Bigl( {\dot {\bar \Delta}^{(0)}} \gamma_\rho {\bar \Delta}^{(0)} \gamma_\sigma   \Bigr)
+ {\rm tr} \Bigl( {\bar \Delta}^{(0)} \gamma_\rho {\dot {\bar \Delta}^{(0)}} \gamma_\sigma   \Bigr)
\Bigr]
\nn\\
&~~= - \frac{1}{2}{\rm tr}\Bigl(  \otimes {\bar\Delta}^{(0)}  \tau_{e^4}^{{\bar \Psi}\Psi} {\bar\Delta}^{(0)}\Bigr)\,.
\label{2nd term b}
\end{align}
Here $\bigl(\tau_{e^4}^{{\bar \Psi}\Psi}\bigr)_{{\hat \alpha} \beta}$
represents a pair of vertices shown in \Cref{fig:tau-e4} obtained from
$\Gamma_{e^4}^{{\bar \Psi}\Psi {\bar \Psi}\Psi}$ in
Figs. \ref{four-Fermi_e4} and \ref{four-Fermi_e4-2}.  The expression in
\labelcref{2nd term b} is a fermion one-loop with the vertex
$\tau_{e^4}^{{\bar \Psi}\Psi}$ and the regulator inserted: it is the
expected one-loop form on the l.h.s. of \labelcref{Condition on
corrections}.  The vertex ${\Gamma}_{\rm e^2}^{\bar\Psi\Psi}$ has a
dressed photon propagator and the scale change also acts on ${\cal
A}_{\mu\nu}$ in the propagator.  Still we have shown that all the terms
of ${\dot\Gamma}_{\rm e^2}^{\bar\Psi\Psi}$ in \labelcref{pa_t
fermi-2-point} are expressed as one-loop forms on the l.h.s. of
\labelcref{Condition on corrections}.

We also studied $\Gamma_{e^3}^{{\bar \Psi} A \Psi}$, which also has a
dressed photon propagator, and confirmed that all the terms in the scale
derivative of the vertex are expressed in one-loop forms on the
l.h.s. of \labelcref{Condition on corrections} by including an
appropriate RG-improved one-loop vertices.

In the above, we have explained that \labelcref{Condition on corrections}
holds for the first three terms in the expansion \labelcref{mod.strlog2-2}.
Though we have not proved that arguments similar to the above applies to
all the terms in the expansion \labelcref{mod.strlog2-2}, the above results
implies strongly that $\Gamma_I (t)$ in \labelcref{Gamma I at t} satisfies
the flow equation with the classical vertices $\tau_{\rm cl}$ and
appropriate RG-improved one-loop vertices.

\subsection{On the dimensionless flow equation}
\label{app:dimless flow}

The flow equation \labelcref{eq:wetterich} in the dimensionless form
collects quantum corrections in a different manner from that explained
above.  To point out the difference is the purpose of this subsection.

By putting an Ansatz on the r.h.s. of the flow equation, we add higher
order quantum corrections that makes the action different from the
original Ansatz.  Using a derivative expansion, we reduces the resultant
action into the form of our Ansatz but with different couplings.  The
renormalization with the Z factors is introduced in this step and we
find the flow equations of couplings.  The process of accumulating
higher order quantum corrections defines the non-perturbative runnings
of couplings though the form of the 1PI action is always kept in the
form of a chosen Ansatz.  This is the approximation for the flow
equation \labelcref{eq:wetterich} to obtain \labelcref{eq:flows}.

\bibliographystyle{plain}

\begin{thebibliography}{10}

\bibitem{Igarashi:2000vf}
Y.  Igarashi, K.  Itoh, and H. So, Prog. Theor. Phys. {\bf 104},
  1053--1066 (2000),  {{arXiv:0006180}}.

\bibitem{Igarashi:2001mf}
Y.  Igarashi, K.  Itoh, and H. So, Prog. Theor. Phys. {\bf 106},
  149--166 (2001),  {{arXiv:0101101}}.

\bibitem{Igarashi:2007fw}
Y.  Igarashi, K.  Itoh, and H. Sonoda, Prog. Theor. Phys. {\bf
  118}, 121--134 (2007),  {{arXiv:0704.2349}}.

\bibitem{Higashi:2007ax}
T. Higashi, E. Itou, and T. Kugo, Prog. Theor. Phys. {\bf 118},
  1115--1125 (2007),  {{arXiv:0709.1522}}.

\bibitem{BATALIN198127}
I. A. Batalin and G. A. Vilkovisky, Phys. Lett. B, {\bf 102}(1), 27--31
  (1981).

\bibitem{Fu:2022gou}
W-j. Fu, Commun. Theor. Phys. {\bf 74}(9), 097304 (2022),
  {{arXiv:2205.00468}}.

\bibitem{Dupuis:2020fhh}
N.~Dupuis, L.~Canet, A.~Eichhorn, W.~Metzner, J.~M. Pawlowski, M.~Tissier, and
  N.~Wschebor, Phys. Rept. {\bf 910}, 1--114 (2021),  {{arXiv:2006.04853}}.

\bibitem{Braun:2011pp}
J. Braun, J. Phys. G {\bf 39}, 033001 (2012),  {{arXiv:1108.4449}}.

\bibitem{Rosten:2010vm}
O.~J. Rosten, Phys. Rept. {\bf 511}, 177--272 (2012),
  {{arXiv:1003.1366}}.

\bibitem{Igarashi:2009tj}
Y.  Igarashi, K.  Itoh, and H. Sonoda, Prog. Theor. Phys. Suppl.
  {\bf 181}, 1--166 (2010),  {{arXiv:0909.0327}}.

\bibitem{Gies:2006wv}
H.  Gies, Lect. Notes Phys. {\bf 852}, 287--348 (2012),
  {{arXiv:hep-ph/0611146}}.

\bibitem{Pawlowski:2005xe}
J.~M. Pawlowski, Ann. Phys. {\bf 322}, 2831--2915 (2007),
  {{arXiv:hep-th/0512261}}.

\bibitem{Bambi:2023jiz}
C. Bambi, L. Modesto, and I. Shapiro, editors,
\newblock {\em {Handbook of Quantum Gravity}},
\newblock  (Springer, 2024).

\bibitem{Pawlowski:2020qer}
J.~M. Pawlowski and M. Reichert, Front. in Phys. {\bf 8}, 551848 (2021),
  {{arXiv:2007.10353}}.

\bibitem{Ihssen:2025cff}
F. Ihssen and J.~M. Pawlowski,  {{arXiv:2503.22638}}.

\bibitem{Falls:2025tid}
K. Falls,  {{arXiv:2503.05869}}.

\bibitem{Bergner:2012nu}
G. Bergner, F. Bruckmann, Y. Echigo, Y. Igarashi, J.~M. Pawlowski,
  and S. Schierenberg, Phys. Rev. D {\bf 87}(9), 094516 (2013),
  {{arXiv:1212.0219}}.

\bibitem{Igarashi:2019gkm}
Y.  Igarashi, K.  Itoh, and T.~R. Morris, Prog. Theor. Exp. Phys. {\bf 2019}(10), 103B01
  (2019),  {{arXiv:1904.08231}}.

\bibitem{Igarashi:2021zml}
Y.  Igarashi and K.  Itoh, Prog. Theor. Exp. Phys. {\bf 2021}(12), 123B06 (2021),
  {{arXiv:2107.14012}}.

\bibitem{Igarashi:2016gcf}
Y.  Igarashi, K.  Itoh, and J.~M. Pawlowski, J. Phys. A {\bf 49}(40),
  405401 (2016),  {{arXiv:1604.08327}}.

\bibitem{Maskawa:1974vs}
T. Maskawa and H. Nakajima, Prog. Theor. Phys. {\bf 52}, 1326--1354
  (1974).

\bibitem{10.1143/PTP.54.860}
T. Maskawa and H. Nakajima, Prog. Theor. Phys. {\bf
  54}(3), 860--877 (09 1975).

\bibitem{Fukuda:1976zb}
R. Fukuda and T. Kugo, Nucl. Phys. B {\bf 117}, 250--264 (1976).

\bibitem{Miransky:1984ef}
V.~A. Miransky, Nuov. Cim. A {\bf 90}, 149--170 (1985).

\bibitem{Kondo:1988qd}
K-I. Kondo, H. Mino, and K. Yamawaki, Phys. Rev. D {\bf 39},
  2430 (1989).

\bibitem{Bardeen:1985sm}
W.~A. Bardeen, C.~N. Leung, and S.~T. Love, Phys. Rev. Lett. {\bf
  56}, 1230 (1986).

\bibitem{LEUNG1986649}
C. N. Leung, S. T. Love, and W.~A. Bardeen, Nucl. Phys. B {\bf 273}(3),
  649--662 (1986).

\bibitem{Aoki:1996fh}
K-I. Aoki, K-I. Morikawa, J-I. Sumi, H. Terao, and M. Tomoyose,
	Prog. Theor. Phys. {\bf 97}, 479--490 (1997),
  {{arXiv:hep-ph/9612459}}.

\bibitem{Gies:2020xuh}
H.  Gies and J. Ziebell, Eur. Phys. J. C {\bf 80}, 607 (2020),
  {{arXiv:2005.07586}}.

\bibitem{Gies:2022aiz}
H.  Gies, Kevin K.~K. Tam, and J. Ziebell, Eur. Phys. J. C {\bf 83}(10),
  955 (2023),  {{arXiv:2210.11927}}.

\bibitem{Bonini:1994kp}
M.~Bonini, M.~D'Attanasio, and G.~Marchesini, Nucl. Phys. {\bf B437}, 163--186
  (1995),  {{arXiv:hep-th/9410138}}.

\bibitem{Ellwanger:1995qf}
U. Ellwanger, M. Hirsch, and A. Weber, Z. Phys. C {\bf 69}, 687--698
  (1996),  {{arXiv:hep-th/9506019}}.

\bibitem{Ellwanger:1996wy}
U. Ellwanger, M. Hirsch, and A. Weber, Eur. Phys. J. C {\bf 1},
  563--578 (1998),  {{hep-ph/9606468}}.

\bibitem{Bergerhoff:1997cv}
B. Bergerhoff and C. Wetterich, Phys. Rev. D {\bf 57}, 1591--1604
  (1998),  {{arXiv:hep-ph/9708425}}.

\bibitem{Mitter:2014wpa}
M. Mitter, J.~M. Pawlowski, and N. Strodthoff, Phys. Rev. D {\bf 91},
  054035 (2015),  {{arXiv:1411.7978}}.

\bibitem{Cyrol:2016tym}
A.~K. Cyrol, L. Fister, M. Mitter, J.~M. Pawlowski, and N.
  Strodthoff, Phys. Rev. D {\bf 94}(5), 054005 (2016),  {{arXiv:1605.01856}}.

\bibitem{Cyrol:2017ewj}
A.~K. Cyrol, M. Mitter, J.~M. Pawlowski, and N. Strodthoff, Phys. Rev.
  D {\bf 97}(5), 054006 (2018),  {{arXiv:1706.06326}}.

\bibitem{Corell:2018yil}
L. Corell, A.~K. Cyrol, M. Mitter, J.~M. Pawlowski, and N.
  Strodthoff, SciPost Phys. {\bf 5}(6), 066 (2018),  {{arXiv:1803.10092}}.

\bibitem{Fu:2025hcm}
W-j. Fu, C. Huang, J.~M. Pawlowski, Y-y. Tan, and L-j. Zhou,  {{arXiv:2502.14388}}.

\bibitem{Wetterich:1992yh}
C. Wetterich, Phys. Lett. B {\bf 301}, 90--94 (1993).

\bibitem{Morris:1993qb}
T.~R. Morris, Int. J. Mod. Phys. A {\bf 9}, 2411--2450 (1994),
  {{hep-ph/9308265}}.

\bibitem{Bonini:1992vh}
M.~Bonini, M.~D'Attanasio, and G.~Marchesini, Nucl. Phys. B {\bf 409}, 441--464
  (1993),  {{hep-th/9301114}}.

\bibitem{Ellwanger:1993mw}
U. Ellwanger, Z. Phys. C {\bf 62}, 503--510 (1994),
  {{arXiv:hep-ph/9308260}}.

\bibitem{PhysRevLett.93.110405}
H.  Gies and J. Jaeckel, Phys. Rev. Lett. {\bf 93}, 110405 (Sep 2004).

\bibitem{Gies:2003dp}
H.  Gies, J. Jaeckel, and C. Wetterich, Phys. Rev. D {\bf 69},
  105008 (2004),  {{hep-ph/0312034}}.

\end{thebibliography}

\end{document}